\documentclass[aps,12pt,final,notitlepage,oneside,onecolumn,nobibnotes,nofootinbib,noshowpacs,centertags]{revtex4}
\usepackage{graphicx}
\usepackage{dcolumn}
\usepackage{bm}
\begin{document}

\title{Nature of the metal-nonmetal transition in metal-ammonia solutions. I.
Solvated electrons at low metal concentrations.}
\author{Gennady N.~Chuev$^{1,3}\footnote{Authors to whom correspondence should be addressed, e-mail address: genchuev@rambler.ru pascal.quemerais@grenoble.cnrs.fr}$, Pascal Qu\'{e}merais$^2$, and Jason Crain$^{3,4}$}
\date{\today}

\begin{abstract}
Using a theory of polarizable fluids, we extend a variational treatment of an excess electron to the many-electron case corresponding to finite
metal concentrations in metal-ammonia solutions (MAS). We evaluate dielectric, optical, and thermodynamical properties of MAS at low metal
concentrations. Our semi-analytical calculations based on a mean-spherical approximation correlate well with the experimental data on the
concentration and temperature dependencies of the dielectric constant and the optical absorption spectrum. The properties are found to be
mainly determined by the induced dipolar interactions between localized solvated electrons, which result in the two main effects: the dispersion
attractions between the electrons and a sharp increase in the static dielectric constant of the solution. The first effect creates a classical
phase separation for the light alkali metal solutes (Li, Na, K) below a critical temperature. The second effect leads to a dielectric
instability, i.e., polarization catastrophe, which is the onset of metallization. The locus of the calculated critical concentrations is in a
good agreement with the experimental phase diagram of Na-NH$_3$ solutions. The proposed mechanism of the metal-nonmetal transition is quite
general and may occur in systems involving self-trapped quantum quasiparticles.
\end{abstract}

\affiliation{$^{(1)}$ Institute of Theoretical and Experimental Biophysics, \\
Russian Academy of Science, Pushchino, Moscow Region, 142290, Russia \\
$^{(2)}$Institut N\'eel, CNRS, BP 166, 38042 Grenoble Cedex 9, France \\
$^{(3)}$ School of Physics, The University of Edinburgh, Mayfield Road, Edinburgh EH9 3JZ, United Kingdom \\
$^{(4)}$ National Physical Laboratory, Hampton Road, Teddington, TW11OLW, United Kingdom}
\maketitle

\section{Introduction \label{sec1}}

Metal-ammonia solutions (MAS) have been the subject of numerous experimental and theoretical studies, since their discovery by Weyl in 1864.
Many experiments have been performed and a large volume of experimental data have been accumulated (for a review, see
\cite{thompson,C1,C2,edwards}). However MAS still remain essentially an unsolved problem. The interest in this fascinating
system is twofold: first, it exhibits a rich variety of different phases including a phase separation and a metal-to-insulator transition, and second, the
the electronic concentration is low in the metallic state so that correlation effects are important. This intriguing system therefore
remains of sustained interest \cite{rw1,rw2,rw3,rw4,rw5,rw6,rw7,rw8,rw9,rw10}.

Once an alkali metal is dissolved in liquid ammonia, it immediately dissociates to give separated entities with unlike charges: the solvated
ions and electrons. At these concentrations, the solvated electrons form a cavities free of solvent due to short-range interactions with ammonia
molecules, in which they localize with the help of the polarization carried by the surrounding ammonia molecules \cite{jortner1}. Path integral
simulations \cite{deng,K2,K3,K4} and density functional approaches \cite{chuev,Bip}, providing an evaluation of the microscopic structure around
solvated electrons, indicate also the possibility of solvated dielectrons, i.e. paired electrons trapped in a cavity, eventually as metastable
states. The solution remains nonmetallic (electrolytic) at these low concentrations and has an intense blue color independent of the type of
alkali metal, while the optical absorption spectrum does not change up to concentrations of about $0.1$ mole percent of metal (MPM) \cite{MPM}.
For metal concentrations exceeding 7 MPM, the MAS is a liquid metal with a typical bronze coloration. The conductivity of MAS reaches that of
liquid mercury at concentrations of about $20$ MPM \cite{thompson}. However, the blue and the bronze phases are macroscopically separated in the
intermediate range of concentrations varying from $1$ to $10 $ MPM, resulting in a miscibility gap below a critical temperature
\cite{crauss,CHIEUX}. Numerous experiments give evidences about a metal-nonmetal (MNM) transition in this range (also called metal-insulator
transition (MIT) in the paper), indicated by the evolution of the conductivity, the Hall resistance, the Hall mobility, and the Knight shift
\cite{thompson,jortner1}. In particular, the conductivity of Na-NH$_3$ solutions increases by three orders of magnitude between $3$ and $6$ MPM
\cite{thompson}.

Previous models have considered the Mott mechanism \cite{sienko} or have involved an association of localized electrons in clusters
\cite{mottRMP} to explain the MNM transition, but were not able to explain satisfactorily the whole phase diagram. Concerning the Mott scenario,
for example, which is the usual MIT mechanism in doped semiconductors, the MAS seem to be an exception because a phase separation
arises in the concentration range where the MNM transition occurs. We have recently shown \cite{comptes} that the MNM transition is rather driven by the
old Goldhammer-Herzfeld (GH) mechanism of metallization \cite{goldhammer,herzfeld}: the quantum fluctuations of the electrons localized in
cavities induce a polarization catastrophe resulting in metallization of the solution. This proposal follows our previous studies of polarons
interacting at low densities, for which a similar scenario may occur \cite{quem1,quem2,quem3}, and even an analogy between high-$T_c$ cuprates
and MAS was also proposed \cite{quem4}.

We emphasized the prominent role of the dipolar-dipolar interactions \cite {comptes}, i.e. van der Waals interactions in the sense of the London
dispersion forces in the MNM transition. Initially, the idea seems unlikely, since these forces are usually very weak with respect to the
Coulomb interactions in ionic systems. In fact, the situation is more complicated. Diluted MAS are simply electrolytes consisting of solvated
cations (metal ions) and solvated electrons playing the role of anions in the solution. The two species have a different behavior, (classical or
quantum), depending on which energy (or time) scale is considered. Both species behave classically at long time scale. The same is true of the
degrees of freedom: the cation coordinates and the center of mass coordinates of each solvated electron may be treated classically. We note
\cite{comptes} that at the experimental densities of the NMM transition ($\sim 4$MPM) the Debye length is estimated to be $1$\AA\ that is lower
than the mean distance between the ions ($\sim 12$\AA) by an order of magnitude. The electrostatic Coulomb interactions are thus considerably
screened and may not be responsible of the MNM transition. Overlaps between the localized electron wave functions also remain negligible at this
concentration. Then the next step is to consider polarizabilities of the different species: solvent, cations, and solvated
electrons, which are respectively given by: $\alpha _{NH_{3}} \sim18.8$ $%
a_0^3 $ \cite{ammoniapolariz}, $\alpha _{Na^{+}} \sim 1.34$ $a_0^3$ \cite{sodiumpolariz} (for sodium metal), $\alpha_0(0) =e^{2}/m\omega
_{0}(T)^{2}\sim 913$ $a_0^3$ (in Born units $a_0=\hbar^2/me^2$, where $m$ and $e$ are respectively the electron mass and charge). The huge
polarizability of the solvated electrons is due to the internal electronic transition in the cavity. It occurs roughly around $\omega_0 \sim
0.9$ eV, leading to the above polarizability. Notice that these electronic transitions are of Franck-Condon type since the relevant time scale
$\omega_0^{-1}$ is much shorter than the relaxation time $\tau$ of the solvent around localized electrons, which is about several picoseconds
\cite {relaxation}. This polarizability of the solvated electrons induces instantaneous forces of two kinds. First, there are dipole-charge
forces: the interaction between the induced dipole momentum of electrons and the classical degrees of freedom. These forces vanish on average
due to the spherically symmetric distribution of charged species. Secondly, it also induces interaction between quantum fluctuating dipoles,
i.e. dispersion interactions. Since the pioneering work of Onsager, it is well-known that these forces are not additive and must be collectively
treated. This theoretical task have been achieved by several groups and formulated as the theory of quantum polarizable fluids
\cite{pratt,Hoye3,chandler1,shweizer,sq1,chen,sq2,sq3}. We extensively use these results in order to evaluate the contribution of the dispersion
forces between the solvated electrons and their physical consequences. Mainly, the dipole-dipole interactions are self-induced cooperatively, in
contrast to to the case of Coulomb forces which always tend to be self-screened. That is why the dipolar forces may induce a polarization
catastrophe at a finite concentration of metal, provoking the onset of metallization. This scenario essentially follows the GH approach except
that it occurs in a system having charges: the solvated ions and electrons.

Applying a semi-phenomenological treatment, we propose the excess free energy to contain three kinds of terms, namely, an additive contribution
of noninteracting electrons, the excess free energy due to the classical degrees of freedom, and the van der Waals part due to the dispersion
interactions. In the paper, we give a complete theory of the nonmetallic phase, including details which were beyond the scope of our previous
publication \cite{comptes}. We will evaluate thermodynamic quantities, dielectric, and spectroscopic anomalies of the solution. Although the
complete theory of the phase diagram requires the consideration of coexistence of the metallic and the nonmetallic phases, we restrict ourselves
here to the nonmetallic phase and focus on the properties of excess electrons in this phase. The metallic states and a crossover regime will be
considered in our next papers \cite{next}. The layout of this paper is the following. In Section \ref{sec2} we describe our model based on a
variational semi-continuum treatment of MAS and accounting nonpolar, electrostatic, and dipolar interactions between localized electrons. In
Section \ref{sec3} we present the results obtained within the framework of our model. The last section is devoted to discussions, future
extensions, and applications. Two appendices contains special topics: derivation of the effective free energy functional and evaluations of the
effective polarizability for interacting electrons. Atomic units are used throughout.

\section{Model and approximations \label{sec2}}

\subsection{General outline of the problem\label{sec2a}}

Upon dissolving the metal, a broad optical absorption appears around $\omega _{0}\sim 0.9$ eV, which does not depend on nature of the solute.
This experimental fact indicates that solvated electrons dissociated from metal ions are responsible for the absorption and form independent
entities \cite {thompson} . The problem of a single solvated electron has been intensively studied (see Sec. \ref{sec2b}), however the many-body
problem posed by a finite concentration of solvated electrons is a more difficult task, because it includes classical and quantum correlations,
as well as short and long-range interactions. Indeed, there are few theoretical approaches to investigate electrons solvated in MAS at finite
metal concentrations except the ones exceeding $10$ MPM, where the system is highly metallic and localized electrons have disappeared
\cite{ashcroft}.

We propose a semi-phenomenological model following which we treat the main relevant physical features of the diluted MAS (sketched on Fig.
\ref{fig0}). As will be discussed later (Sec. \ref{sec2b}), the single solvated electron is localized in a cavity of diameter $\sigma $ and
surrounded by solvent molecules. The number of molecules which compose the solvated cavity is estimated to be about of 8 in diluted solutions
($<5$ MPM) \cite {thompson}. The individual cavity diffuses as a whole in the solution due to the thermal fluctuations exactly as
would do a classical negatively charged ion. The variables $\{\mathbf{R}%
_{-}^{\{N\}}\}$ determine their cartesian coordinates, while $\{\mathbf{R} _{+}^{\{N\}}\}$ are the cartesian coordinates of the solvated
cations. For sake of simplicity, we assume that each solvated cation and each solvated electron have the same diameter $\sigma $. Thus, from an
electrostatic point of view, the diluted MAS behaves as an electrolyte consisting of cations and anions with diameters $\sigma $ dissolved in a
solvent with dielectric constant $\epsilon (\omega =0)=\epsilon _{s}$, which is the one of pure liquid ammonia. Such a solution may be
statistically treated by the restricted primitive model (RPM), while an analytical expression of the free energy can be found by within the RPM
(see Sec. \ref{sec2c}).

But this is not a complete description (see Appendix 1). In addition to this classical part, there are extra degrees of freedom for the solvated
electrons due to the possible individual electronic transitions inside their solvated cavities. We introduce variables $\{ \mathbf{u}^{\{N\}}
\}$ which are the relative cartesian coordinates of the electrons with respect to their centers of mass $\{\mathbf{R}_{-}^{\{N\}}\}$. The
degrees $\{ \mathbf{u}^{\{N\}}\}$ are quantum mechanical in contrast to the variables $ \{\mathbf{R}_{-}^{\{N\}}\}$. This separation between the quantum and
the classical degrees of freedom may be viewed as a Born-Oppenheimer approximation justified by the relevant energy scales, i.e. $k_{B}T$ for the
classical degrees and $\hbar \omega _{0}$ for the quantum ones, which magnitudes roughly differ by two orders. Physically, it means that the
duration of an electron transition from its local ground-state to an excited state ($\sim \omega _{0}^{-1}$) is sufficiently small with respect
to the relaxation time of the solvent. Consequently, the quantum correlations or interactions induced by these electronic transitions are
screened by the high-frequency dielectric constant of the solvent, i.e. $\epsilon (\omega \sim \infty )=\epsilon _{\infty }$. This difference of
screening for the
different interactions, $\epsilon _{s} \sim 20$ for the classical ones, and $%
\epsilon _{\infty } \sim 1.7$ for the quantum ones, plays a significant role in our theory as it was already identified for the case of the
melting of a polaron Wigner crystal \cite{quem1,quem2,quem3}. We carry out in Appendix 1 an expansion of Coulomb interactions and show that it
includes the dipole-dipole interactions between the solvated electrons apart from the classical electrostatic interactions.  This is nothing but
their dispersion interactions as already said in the Introduction.  They play a dominant role in the thermodynamical behavior of diluted MAS
(see Sec. \ref{sec3b}) owing to the conjugated effects of the large polarizability of solvated electrons and the weak screening by the solvent
($\epsilon _{\infty }$ as we have seen).
Therefore at metal density $n<5$MPM, the change $%
\Delta f_{nm}(n)$ in the total excess free energy per solvated electron may be written as the sum of three terms:
\begin{equation}
\Delta f_{nm}(n)=f_{0}+\Delta f_{cl}(n)+\Delta f_{d}(n),  \label{total}
\end{equation}
where $f_{0}$ is excess free energy of a single electron, $\Delta f_{cl}(n)$ is the contribution of the classical degrees of freedom, and
$\Delta f_{d}(n)$ is the van der Waals contribution due to dispersion forces between solvated electrons.

\subsection{Free energy of a single solvated electron\label{sec2b}}

Numerous theoretical and experimental studies \cite
{thompson,jortner1,deng,K3,chuev,UFN} have indicated that the polarization
due to the surrounding molecules and the cavity formation due to short range
repulsion plays the main role in the electron solvation. Following this
concept, the free energy of an excess electron in an infinitely diluted MAS
may be written as
\begin{eqnarray}
f_{0}(r_{e}) &=& \frac{p_{e}^{2}(r_{e})}{2}-[\frac{1}{\varepsilon _{\infty }}%
- \frac{1}{\varepsilon _{s}}]\frac{1}{2r_{e}} + 4\pi C_{S}\lambda r_{e}^{2}+
\frac{4\pi C_{V}n_{s}}{3\beta }r_{e}^{3},  \label{1}
\end{eqnarray}
where the first term $p_{e}^{2}/2$ is the quantum mechanical kinetic energy of localization for
the solvated electron. $p_{e}$ is the mean momentum of the electron which
depends on the cavity radius $r_{e}$. The Heisenberg inequality leads to $%
p_{e}\approx C_{k}/r_{e}$, where $C_{k}$ is a constant of proportionality which depends on details of the electron density distribution. The
second term in (\ref{1}) is the potential contribution due to the polarization of solvent. This polarization has two contributions due to the
existence of two sources of polarization (as in the polaron problem). The first depends on the electronic polarizability of the solvent
molecules (ammonia) and provides the high frequency dielectric response $\varepsilon _{\infty }$. The second corresponds to the orientational
polarization of the ammonia molecules. It provides, together with the first, the low frequency (static) dielectric response $\varepsilon _{s}$
of the solvent. The last two phenomenological terms in (\ref{1}) are the nonpolar contributions due to the cavity formation, i.e. the work
necessary to perform the cavity formation. It contains a surface term ($\lambda $ is the surface tension) and a volume contribution
corresponding to the work of an external pressure $p_{s}\propto \beta ^{-1}n_{s}$. It depends on the solvent density $n_{s}$, temperature $\beta
^{-1}$, and details of the local structure around the cavity formation (the size of the first coordination shell, the number and the orientation
of solvent molecules in the shell, etc). We approximate the influence of all the details by the two phenomenological parameters $C_{S}$ and
$C_{V}$, which can be evaluated within the microscopic theories \cite{25,c4}, or
adjust by fitting experimental data. The extremum of the free-energy $%
f_{0}(r_{e})$ provides consistent evaluations of the cavity radius $r_{e}$, i.e. $\partial f_{0}(r_{e})/\partial r_{e}=0$. Equation (\ref{1}) is
a modified variational treatment of the one-polaron problem \cite{pekar} within the framework of the semi-continuum approach \cite{jortner1}.

To take into account the internal electronic transition occurring at
frequency $\omega_0(T)$, we relate phenomenologically this frequency to the
cavity size as
\begin{equation}
\omega _{0}(T)=\frac{C_{r}(T)}{r_{e}^{2}},  \label{omega}
\end{equation}
where $C_{r}(T)$ is a parameter depending both on the temperature and on the local potential felt by the internal degrees  $\{ \mathbf{u}_i\}$.
 It is well-known that $C_{r}=3/2$ for a three-dimensional parabolic potential, however  the constant deviates from
this value for anharmonic oscillators. The simple analysis of experimental data yields $C_{r}\approx 1.25$ for the infinitely diluted MAS at
$T=-70^{0}C$ \cite{jpc95}, taking $r_e=3.2$ \AA\ and $\omega_0=0.9$ eV. This indicates weak deviations of the actual potential from the harmonic
oscillator. Hence, the constant $C_{r}(T)$ can be fitted to provide relations between the calculated values of $r_{e}$ and the experimental data
on $\omega _{0}(T)$.

\subsection{Classical contribution to the excess free energy\label{sec2c}}

The classical part of the excess free energy is composed by charged hard spheres of same diameter $\sigma$ dissolved in a solvent with
dielectric constant $\epsilon_s$. It yields two contributions: the nonpolar ($\Delta f_{n}(n)$) and the coulomb ($\Delta f_c(n)$) parts:
\begin{equation}
\Delta f_{cl}(n)=\Delta f_{n}(n) + \Delta f_{c}(n).
\end{equation}
Analytical expressions for these contributions are well-known. Following the Carnahan-Starling result \cite{CS,RMP}, the nonpolar part is
expressed as:
\begin{equation}
\beta \Delta f_{n}(n) = 2 \left[ \ln (n/n_s)-1+\frac{\eta (4-3\eta )}{(1-\eta )^{2}}\right], \label{fhs}
\end{equation}
where $n_s$ is the solvent density, while $\eta = \pi n\sigma ^{3}/3$ is the packing factor. In the above expression, we take into account that
the density of hard spheres is equal to $2n$ (cations plus solvated electrons). The first term is the ideal part, while the last term is the
hard-sphere exclusion part. Since for the densities that we consider $\eta << 1$, the ideal term is preponderant. For example, at $n=4$ MPM,
$\eta \sim 0.04$, and only few terms in the low-density expansion of (\ref{fhs}) are physically relevant.

For the electrostatic contribution, we use the results of the mean spherical approximation (MSA). The contribution $\Delta f_{c}(n)$ is
expressed in terms of the inverse screening length \ $\gamma =([1+2\sigma /\ell_D ]^{1/2}-1)/2\sigma $, where the Debye length is $\ell_D= (8\pi
\beta n e^2/\epsilon _{s})^{-1/2}$. The Coulomb free energy part is then given by \cite{blum}:
\begin{equation}
\beta \Delta f_{c}(n)=-\frac{2\beta }{\epsilon _{s}[\sigma +\gamma ^{-1}]}+%
\frac{\gamma ^{3}}{3\pi n}.
\end{equation}

\subsection{Van der Waals contribution to the excess free energy \label{sec2d}}

The most difficult part of the free energy to evaluate, comes from the contribution of the dispersion forces between solvated electrons. Because
we have separated the quantum and the classical variables, we may write the quantum part as the hamiltonian of interacting Drude oscillators
(Appendix 1). Introducing the dipolar momentum ${\bf m}_i=e {\bf u}_i$, the quantum part of the Hamiltonian for a given configuration $\{ {\bf
R}_-^{\{ N \}}\}$ is written as:
\begin{eqnarray}
H=\sum_{i=1}^N \frac{\alpha_0 \omega_0^2 \pi_i^2}{2} + \frac{m_i^2\alpha_0}{2}+ \frac {1}{2} \sum_{i \ne j} u_{hs}(\vert {\bf R}_-^i-{\bf R}_-^j
\vert)+ \frac{1} {2 \epsilon_{\infty}} \sum_{i \ne j}^N {{\bf m}_i \cdot {\bf T}({\bf R}^i_--{\bf R}^j_-) \cdot {\bf m}_j} , \label{drude-osci}
\end{eqnarray}
where ${\mathbf \pi}_i$ is the conjugate of the variable ${\bf m}_i$, and $\alpha_0 = e^2/m \omega_0^2$ is the bare polarizability of a solvated
electron, $u_{hs}(r)$ is the hard-sphere interaction potential $u_{hs}(r> \sigma)=0$ and $u_{hs}(r<\sigma)=\infty$, while ${\bf T}( {\bf r})=3
r^{-5} {\bf r}{\bf r}-r^{-3} {\bf I}$ is the dipolar rank-2 tensor.

To evaluate different physical properties, two quantities are important: the density of states per particle $D(\omega, \{ {\bf R}_-^{\{ N
\}}\})$, and the average polarizability $ \alpha (\omega,\{ {\bf R}_-^{\{ N \}}\} )$. Both are related to the eigenvalues $ \{ \omega_{\lambda}
(\{ {\bf R}_-^{\{ N \}}\} ) \}$ of the {\it collective modes} of (\ref{drude-osci}) and depend on the given configuration $\{ {\bf R}_-^{\{ N
\}}\}$:
\begin{eqnarray}
D(\omega, \{ {\bf R}_-^{\{ N \}}\}) &=& \frac{1}{3N} \sum_{\lambda=1}^{3N} \delta (\omega-\omega_{\lambda}(\{ {\bf R}_-^{\{ N \}}\})), \nonumber
\\ \alpha (\omega, \{ {\bf R}_-^{\{ N \}}\}) &\sim& \frac {1}{3N} \sum_{\lambda=1}^{3N} \frac{1}{\omega_{\lambda}(\{ {\bf R}_-^{\{ N \}}\})^2-
\omega^2-i \Gamma_{\lambda}}.
\end{eqnarray}
Using mathematical relations of the theory of distributions, namely, $Im(1 / (\omega-i0)) = \pi \delta(\omega)$ and $\delta(\omega^2-\omega_0^2)
= \delta(\omega-\omega_0)/2\omega$, we can directly relate the above quantities:
\begin{equation}
D(\omega, \{ {\bf R}_-^{\{ N \}}\}) = \frac{6 \omega \alpha_i(\omega, \{ {\bf R}_-^{\{ N \}}\})}{\pi},
\end{equation}
where $\alpha_i(\omega,\{ {\bf R}_-^{\{ N \}}\})$ is the imaginary part of $\alpha(\omega,\{ {\bf R}_-^{\{ N \}}\})$.

The next step is to proceed the thermodynamical average over the classical degrees of freedom in order to define the DOS and the effective
polarizability of a solvated electron:
\begin{eqnarray}
D(\omega) =\left\langle D(\omega, \{ {\bf R}_-^{\{ N \}}\}) \right\rangle,\qquad \alpha(\omega) &=& \left\langle \alpha (\omega, \{ {\bf
R}_-^{\{ N \}}\}) \right\rangle , \label{d-alpha}
\end{eqnarray}
where the brackets $\left\langle \cdots \right\rangle$ denotes the thermodynamical average over the variables $\{ {\bf R}_-^{\{ N \}}\}$.
Rigourously, the thermodynamical average should also include average over the positive ions ($\{ {\bf R}_+^{\{ N \}}\}$), taking into account
the correlations between classical particles due to the Coulomb interactions. However, as we have discussed in the Introduction, the Debye
length is very small ($\sim 1$\AA\ ) at the densities that we are considering, so that we neglect the correlations due to the Coulomb
interactions in the thermodynamical average, taking only into account the correlations due to the dipolar and the nonpolar interactions. Under
this assumption, we are left with a polarizable neutral fluid represented by (\ref{drude-osci}), in which the classical variables $\{ {\bf
R}_-^{\{ N \}}\}$ now implicitly correspond to the positions of {\it neutral} classical hard spheres.

In the framework of the theory of polarizable fluids, the problem has been extensively studied
\cite{pratt,Hoye3,chandler1,shweizer,chen,sq1,sq2,sq3}, we use this approach and refer the reader to these references for more details. The main result is
that the effective polarizability $\alpha (\omega )$ defined by (\ref{d-alpha}) is calculated via the self-consistent equation:
\begin{equation}
\alpha ^{-1}\left( \omega \right) =\alpha _{0}^{-1}\left( \omega \right)
-2U(\alpha \left( \omega \right) /\epsilon _{\infty })/\epsilon _{\infty },
\label{effectivealpha}
\end{equation}
where the last term accounts for the correlations between induced dipoles,while $\alpha_0(\omega) \sim (\omega_0^2-\omega^2)^{-1}$ is the bare
frequency polarizability of the particles. Equation (\ref{effectivealpha}) has been derived in \cite{chandler1}, we only modify it by taking
into account the high-frequency screening by the solvent (the use of $\epsilon _{\infty }$ in (\ref{effectivealpha})). The function $3\alpha
U(\alpha )/\beta $ corresponds to the dipolar part of the internal energy per particle of a classical liquid of nonpolarizable particles with
permanent dipole momentum $(3\alpha /\beta )^{1/2}$ \cite {chandler1,pratt}. Once this function is known, all the physical properties of the
system can be evaluated by solution (\ref{effectivealpha}). The simplest method to obtain $U(\alpha )$ is the Pad\'{e} approximation \cite
{chandler1,shweizer}, which is an interpolation between the case of low and large polarizabilities $\alpha (\omega )$ (Appendix 2). Adapting
this method to our case, we get:
\begin{eqnarray}
U(\alpha (\omega )/\epsilon _{\infty })=\frac{I_{0}(nr_{e}^{3})n\alpha
(\omega )}{8\epsilon _{\infty }r_{e}^{3}+I_{1}(nr_{e}^{3})\alpha (\omega )},
\label{pade}
\end{eqnarray}
where $I_{0}(x)$ and $I_{1}(x)$ are analytical functions depending on dimensionless density $x=8nr_{e}^{3}$ \cite{pratt}. As a result, we obtain
the quadratic algebraic equation for $\alpha (\omega )$:
\begin{eqnarray}
[\epsilon _{\infty }I_{1}(x)-2I_{0}(x)n\alpha _{0}\left( \omega \right) ]\alpha ^{2}\left( \omega \right)    =[8\epsilon _{\infty
}^{2}r_{e}^{3}-\epsilon _{\infty }I_{1}(x)\alpha _{0}(\omega )]\alpha \left( \omega \right) -8\epsilon _{\infty }^{2}r_{e}^{3}\alpha _{0}(\omega
),  \label{quadra}
\end{eqnarray}
whose the complex solution is:
\begin{eqnarray}
\alpha (\omega )=\alpha _{r}(\omega )+i\alpha _{i}(\omega ).
\end{eqnarray}
The imaginary part $\alpha _{i}(\omega )$ is nonzero only in a finite range of frequency $\omega _{-}(T,n)<\omega <\omega _{+}(T,n)$. This is a
direct consequence of the Born-Oppenheimer approximation made in the model (separation between classical and quantum degrees of freedom, as it
is also done in the theory of polarizable fluids). In general, all quantities may be evaluated numerically by the above equations. However, all
the quantities may be found analytically in the special case of dilute solutions for which $nr_{e}^{3}\ll 1$ (see Appendix 2). For example, the
imaginary part of the effective polarizability is given by:
\begin{eqnarray}
\alpha _{i}(\omega )=\frac{3\varepsilon _{\infty }^{2}r_{e}^{3}(\omega _{+}^{2}(n)-\omega ^{2})^{1/2}(\omega ^{2}-\omega
_{-}^{2}(n))^{1/2}}{2\pi n}, \label{alil}
\end{eqnarray}
where $\omega _{\pm }(n)$ are the edge frequencies:
\begin{eqnarray}
\omega _{\pm }=[\omega _{0}^{2}\pm \sqrt{\frac{4\pi n_{e}}{3\varepsilon
_{\infty }^{2}r_{e}^{3}}}]^{1/2}.  \label{edgl}
\end{eqnarray}
In that case the DOS ($D(\omega) \sim \omega \alpha_i(\omega)$)
has a semielliptic form.

Once the above calculations have been performed, it is easy to evaluate the van der Waals contribution to (\ref{total}), which is directly given
by the difference between the free energy of the {\it interacting} Drude oscillators and the free energy of the same but {\it non-interacting}
oscillators \cite{lundq}. For the temperature of interest $T<250$ K and the bare frequency of the dipoles $\omega_0 \sim 1$ eV, we have $\beta
\omega_0 \gg 1$. Then, the free energy per oscillator in the non-interacting case is just the zero-point energy $3\hbar \omega_0/2$. The free
energy is related to the DOS so that the van der Waals contribution is just given by:
\begin{eqnarray}
\Delta f_{d}=\frac{1}{\beta }\int_{0}^{\infty }D(\omega)\ln (2\sinh [\frac{\beta \hbar \omega}{2}])d\omega -\frac{3\hbar \omega _{0}}{2},
\label{vdwE}
\end{eqnarray}
This equation must be evaluated numerically together with $\alpha(\omega)$ and $D(\omega)$. However in the diluted solutions, we may obtain an
analytic expression (Appendix 2):
\begin{equation}
\Delta f_{d}=-\omega _{0}[\frac{\pi n\alpha _{0}^{2}}{16\varepsilon _{\infty }^{2}r_{e}^{3}}+O(n^{2}\alpha _{0}^{4}r_{e}^{-6})].  \label{disp}
\end{equation}
This contribution is negative as it should be and the total free energy $F_{d} \sim n \Delta f_{d}$ is proportional to $n^2$, as in the case of
the van der Waals model for the liquid-gas transition.

\subsection{Dielectric and thermodynamical properties\label{sec2e}}

To make the link with the dielectric properties, we calculate the total dielectric constant of the solution as the one of a set of independent
spherical particles with effective polarizability $\alpha(\omega)$ in a medium with dielectric constant $\epsilon_{NH_3}(T, \omega)$ (pure
ammonia). The total polarization of the sample under the macroscopic field $\bf{E}(\omega)$ is by definition ${\bf P}_{tot}(\omega) =
[(\epsilon(T, \omega)-1)/4 \pi] {\bf E}(\omega)$ where $\epsilon(T, \omega)$ is the requested dielectric constant of the solution. The
polarization may be written as the sum of two contributions: ${\bf P}_{tot}(\omega) = {\bf P}_{NH_3}(\omega) + \Delta{\bf P}(\omega)$ due to the
solvent and the polarizable spheres, respectively. If we assume that the dielectric properties of the solvent in itself are not affected by the
presence of the polarizable spheres, we may write ${\bf P}_{NH_3}(\omega)= [(\epsilon_{NH_3}(T, \omega)-1)/4 \pi ] {\bf E}(\omega)$, while the
dipolar contribution to the polarization is just given by:
\begin{equation}
\Delta {\bf P}(\omega) = \frac{(\epsilon(T, \omega) - \epsilon_{NH_3}(T, \omega))}{4 \pi} {\bf E}(\omega). \label{Pdip1}
\end{equation}
At the same time, the excess polarization due to the quantum polarizable spheres in the solvent is expressed as (see Appendix 2):
\begin{equation}
\Delta{\bf P}(\omega) =n \alpha(\omega) {\bf E}_{\ell}(\omega) / \epsilon_{\infty},
\label{Pdip2}
\end{equation}
where we have assumed that only the high dielectric constant of the solvent screens the polarizability of the spheres (due to their quantum
nature), and ${\bf E}_{\ell} (\omega)$ is the local field viewed by the given set of polarizable spheres. The Lorentz local field expression
gives:
\begin{eqnarray}
{\bf E}_{\ell} (\omega)= {\bf E}(\omega) + \frac{4 \pi {\bf P}_{tot}(\omega)}{3}, \qquad {\bf E}_{\ell} (\omega) =
\frac{(\epsilon_{NH_3}(T,\omega)+2)}{3} {\bf E}(\omega)+ \frac{4 \pi \Delta {\bf P}(\omega)}{3}. \label{localF}
\end{eqnarray}

Using (\ref{Pdip1}), (\ref{Pdip2}), and (\ref{localF}), we obtain the modified Maxwell-Garnett expression \cite{MG}):
\begin{eqnarray}
\frac{\epsilon (T,\omega )-\epsilon _{NH_{3}}(T,\omega )}{\epsilon (T,\omega
)+2\epsilon _{NH_{3}}(T,\omega )}=\frac{4\pi }{3\epsilon _{\infty }}n\alpha
(\omega),  \label{Cl-Mo}
\end{eqnarray}
which gives by inversion the expression of the dielectric constant of the solution as function of its constituents, i.e. the solvent and the
solvated electrons. Notice that in the expression of the total polarization we have ignored the polarization due to the presence of the
classical mobile charges, since it is not vanishing only at very low frequencies due to the very low mobility of the ions in the electrolyte.
Our expression (\ref{Cl-Mo}) is thus valid provided $\omega > \tau^{-1}_{D}$, where typically $\tau_D \sim 10^{-12}s$ \cite{chazalviel}.

Concerning the thermodynamic properties, we may calculate the excess pressure, the excess chemical potential, and the excess compressibility
$\kappa$ due to the presence of the solvated electrons in the solutions by the usual relations:
\begin{equation}
\Delta p=n^{2}\frac{\partial \Delta f_{nm}}{\partial n},\qquad \Delta \mu =\Delta f_{nm}+n\frac{
\partial \Delta f_{nm}}{\partial n},\qquad  \kappa^{-1} = n \frac{\partial \Delta p}{\partial n}. \label{press}
\end{equation}
To complete the calculations, we also mention that the solvated radius $r_e$ may be self-consistently obtained by minimizing the excess free
energy (\ref{total}):
\begin{equation}
\frac{ \partial \Delta f_{nm}(n)}{\partial r_e}=0.
\label{self}
\end{equation}

\subsection{The use of experimental data to fit phenomenological parameters.
\label{sec2f}}

Our input phenomenological parameters are $C_{S}$, $C_{V}$, $C_{k}$, $%
C_{r}(T)$, $\sigma$, $\epsilon _{\infty }$, and $\epsilon _{s}(T)$. The
first four constants can be derived from the experimental data on the
absorption spectrum $\nu (\omega )$ and on the excess molar volume directly related to $
r_{e}$. The mean momentum $p_{e}$ is also related with the first moment of the
absorption spectrum, i.e., $p_{e}=\omega _{1}=\int \omega \nu (\omega
)d\omega $.

In general, the dielectric and thermodynamical properties of the localized electrons should be calculated self-consistently, but it leads to
complicated concentration and temperature dependencies of the cavity size $ r_{e}(n)$ and of the maximum $\omega _{0}(n)$ of the absorption
spectrum. To avoid these cumbersome calculations, we fix these quantities $ r_{e}(n)=r_{e}$ (obtained for $n \rightarrow 0$) and $\omega
_{0}(n)=\omega _{0}$ (also obtained for $n \rightarrow 0$) at the first step of our calculations. Using experimental data on $\nu (\omega )$ and
$ r_{e}$ at zero metal concentration and $T=-70^{0}$C \cite{jpc95}, we obtain $ \ C_{k}\approx 1.5$ and $C_{r}\approx 1.25$, while we set
$C_{S}=1$ and $ C_{V}=1.75$ to obtain the experimental value $r_{e}=3.2$ \AA\ and $\omega _{0}(T=-70^{o}\text{C})=0.9$ eV.

We also use the experimental data on $ \omega _{0}(T)$ \cite{omegaT}. This last parameter leads to an implicit temperature dependence of $\alpha
(\omega )$, we take $\partial \omega _{0}(T)/\partial T=-2.2\cdot 10^{-3}$ eV/K in the temperature range ($ -70^{o}\text{C}<T<+70^{o}\text{C}$)
\cite{omegaT} and use the appropriate data on dielectric constants of liquid ammonia, i.e., $\epsilon _{\infty }=1.756$, $\epsilon
_{s}(T=-70^{0}$C$)=25$, and $\partial \epsilon _{s}/\partial T=-0.1\ $K$^{-1}$ \cite{epsT}. Finally we also need data on the density of pure
ammonia, deriving them from \cite{rhoT}, namely, $n_{NH_{3}}=0.0255$\AA $^{-3}$ at $T=-70^{0}$C, while $\partial n_{NH_{3}}/\partial T=-5\cdot
10^{-5}$ \AA $^{-3}$K$^{-1}$.

\section{Results and analysis \label{sec3}}

\subsection{Criticality of the nonmetallic phase}

Several transitions  can arise in nonmetallic MAS due to a variety of classical and quantum effects. First, we consider quantum instabilities:
one of them is related to the GH scenario of metallization \cite{herzfeld}, following which Herzfeld took the point of divergency of the
dielectric constant as the onset of metallization. Such an instability also occurs for the same reason in our model, and we call it as
$n_{\infty}(T)$. It depends on $T$ due to the temperature dependence of the bare polarizability through the parameter $\omega_0(T)$. Another
quantum instability arises due to softening of the collectives modes, that induce a quantum transition at a critical density $n_{c1}(T)$ also
depending on temperature. We will see in Sec. \ref{sec3a} that both instabilities $n_{\infty}(T)$ and $n_{c1}(T)$ physically coincide and
correspond to the onset of metallization.

Another important consequence of the model is a thermodynamical instability. It is well-known for a classical fluid that the van der Waals
interactions compete with the short-range interactions to drive a liquid-gas phase separation below a critical temperature $T_c^{\ell g}$. The
same transition occurs in our model. Looking at the locus of this "liquid-gas" phase-separation in its low-density regime, the spinodal curve
$n_{s}(T)$ is evaluated and discussed in Sec. \ref{sec3c} by examining (\ref{total}). The term "liquid-gas" refers, in our model, to a phase
separation between a phase with a low density of solvated electrons, and a phase with a higher density of the same objects. In fact, this
classical phase separation is hidden by the quantum MIT at $n_{c1}>n_{s}(T)$ and should be replaced by a phase separation between the insulating
phase of solvated electrons and the metallic phase of free electrons. This point has been already partly discussed by us \cite{comptes}, details
will be reported in our future publications \cite{next}. We only consider here the low density part of the spinodal curve $n_{s}(T)$.

\subsection{Density of states and stability of the insulating phase\label{sec3a}}

The effective polarizability $\alpha (\omega )$ and the DOS $D(\omega)$ can be immediately deduced by (\ref{ri}) and (\ref{d-alpha}) for the
known (and fixed) cavity size $r_{e}=3.2$ \AA\ . The typical behavior of $D(\omega)$ is shown in Fig. \ref{fig1} at various concentrations and
$T=-70^{o}$C, while the dimensionless imaginary part $\alpha _{i}(\omega )/\alpha _{0}(0)$ and real part $ \alpha _{r}(\omega )/\alpha _{0}(0)$
of the effective polarizability are shown in Fig. \ref{fig2} under the same conditions. As it is seen, for a given temperature, the DOS and the
effective polarizability broaden progressively as the concentration $ n$ rises. Notice that the maximum of the imaginary part of
$\alpha(\omega)$ shifts to low frequencies as $n$ increases, while the maximum of the DOS remains located at the same position. At low enough
concentrations the deviations of the DOS and the imaginary part of $\alpha(\omega)$ from a semielliptic forms are minor, but the edge
frequencies $\omega _{\pm }$, given by (\ref{edgl}), monotonically vary with $n$ (see Fig. \ref{fig3}). The DOS broadening is roughly
proportional to $\Delta \omega _{\pm }=\omega _{+}-\omega _{-} \sim n{}^{1/2}$ at low concentrations. More precisely, the low edge $\omega_-(n)$
progressively decreases as $n$ increases, indicating a softening of some collective modes of the solvated electrons. This softening with
increasing density is directly related to the dipolar interactions \cite{bagchi,quem1,quem2}. Then, the natural condition of stability is to be
$\omega_-(n)^2>0$, since the eigenvalues of the collective modes must be real and positive. The system becomes instable when the softening is
achieved at the critical density $n_{c1}$:
\begin{equation}
\omega _{-}(T,n_{c1})=0.  \label{polcat}
\end{equation}
Taking into account the temperature variation of $\omega_0(T)$ as explained in Sec. \ref{sec2f} and calculating self-consistently radius $r_e$
by (\ref{self}), we find that $n_{c1}(T)$ varies from 2 to 6 MPM depending on temperature.  The result is only different slightly from that
obtained in \cite{comptes} (see, Fig. (\ref{fig4})), since we did not use (\ref{self}) in \cite{comptes}.

Rigourously, such an instability may be interpreted in two ways: either some electrons escape from their cavities at $n_{c1}(T)$ and that
corresponds to the onset of metallization; either the system of solvated electrons becomes instable with respect to the formation of a state
having a finite (non-vanishing) permanent dipole momentum $\left\langle {\bf m}(\omega=0) \right\rangle \ne 0$. This possible phase was first
recognized by Turkevich and Cohen \cite {jpcTC,prlTC} as an excitonic insulator (EI), and later studied by different authors
\cite{jcpL,PM,jcpXS,jpcmL1,jpcmL2,prbW}. However, these studies treated neutral atoms, which are completely different from the present case (in
particular, an eventual permanent momentum for the solvated electrons would necessitate a complete reorganization of ammonia molecules around
the cavities).

Regarding the onset of metallization within the framework of a generalized GH criterion, we calculate the dielectric constant of the system by
(\ref{Cl-Mo}) and find that the real part of the dielectric constant diverges (zero denominator) when:
\begin{equation}
\frac{4\pi n_{\infty }(T)\alpha _{r}(0)}{3\varepsilon _{\infty }}=1,
\label{H-criterion}
\end{equation}
which defines the second critical density $n_{\infty}(T)$, i.e. the locus of the polarization catastrophe. The similar criterion was considered
by Herzfeld \cite{herzfeld}, except that the polarizability $\alpha_0$ of \textit{non-interacting} particles was considered. By comparison,
using the bare polarizability in our formula, we obtain $n_{\infty}$ to be three times larger than that calculated with the use of  the effective
polarizability $\alpha_{r}$, (we find 14 MPM instead of 5 MPM at $T=-70^\circ$C). This clearly indicates the fundamental role played by the
induced dipole-dipole interactions in the MNM transition. The comparison between $n_{c1}(T)$ and $n_{\infty}(T)$ shown in Fig.(\ref{fig4})
indicate that they are quite close, i.e.  $n_{c1}(T) \approx n_{\infty}(T)$ and corresponds to the onset of metallization.

\subsection{Dielectric function and optical properties\label{sec3b}}

As explained above, using Eqn  (\ref{Cl-Mo}) we can extract the dielectric function of the solution. An example of the calculated $\epsilon
(\omega )$ is presented in Fig. \ref{fig5}. As it is seen, the characteristic frequencies range having a nonzero imaginary part and a strong
augmentation of the real part of the dielectric function broaden as the concentration rises. At the same time, the maximal values of imaginary
part $\epsilon _{i}(\omega )$ and the real part $\epsilon _{r}(\omega )$ enormously increase with the concentration, and shift to lower
frequencies up the the point of polarization catastrophe at which $\epsilon_r(\omega)$ diverges at $\omega=0$. Beyond this point, i.e. for $n >
n_{\infty} \approx 5$ MPM, the calculations are meaningless since the system is metallic. The calculated real part of the static dielectric
constant may be also compared with experimental data extracted from the literature \cite{pk,pk1}. The comparison is shown in Fig. \ref{fig6}. We
find a good agreement between the theory and the experiments, although slight deviations are observed at high temperatures, probably due to the
presence of metallic electrons ignored in the present study.

Using the obtained data on the frequency-dependent dielectric function $
\epsilon (T,\omega )$, we may also calculate the optical absorption coefficient $A(\omega )$ as:
\begin{equation}
A(\omega )=A_{0}\omega \lbrack \sqrt{\epsilon _{r}^{2}(\omega )+\epsilon
_{i}^{2}(\omega )}-\epsilon _{r}(\omega )]^{1/2},
\end{equation}
where $A_{0}$ is a constant of the proportionality, while $\epsilon _{r}(\omega )$ and $\epsilon _{i}(\omega )$ are respectively the real and
imaginary parts of the dielectric function. Figure \ref{fig7} indicates the calculated concentration dependence of $A(\omega )$ at the locus of
its maximum $\omega _{\max }$ at $T=-65^{0}$C. This maximum decreases at increased metal concentrations, indicating a red shift of the
absorption maximum observable in the experiments \cite{ccc,ccc1}. A slight deviation between the theory and the experiments is however observed.
Thus, this red shift give additional evidence in favor of the GH scenario, because it is a signature of the polarization catastrophe
\cite{quem1}.

\subsection{Thermodynamical properties\label{sec3c}}

We examine the van der Waals part $\Delta f_{d}$. At low concentrations the obtained value is close to our estimation (\ref{disp}), although it
pronouncedly deviates from it for larger concentrations (see Fig. \ref{fig8}). The first and the second order corrections are significant, which
clearly indicates the many-body character of the dispersion contribution. We have also calculated all the contributions of the total excess free
energy, which are depicted in Fig. \ref{fig9}. As it is seen, the total change in the free energy is about $10/\beta $ and weakly varies with
the concentration: the variation does not exceed $\beta ^{-1}$ in the range $0.5MPM<n<5$ $MPM$. The situation is similar with the electrostatic
contribution $\Delta f_{c},$ which only provides an almost constant amount of about $-2/\epsilon _{s}\sigma$ to the free energy. At the same
time, the van der Waals and the nonpolar contributions (essentially the ideal term in (5)) significatively vary, and their delicate balance
results in a nonmonotonic behavior of the excess chemical potential $\Delta \mu_{tot} =\Delta \mu _{d}+\Delta \mu _{c}+\Delta \mu _{n}$ and the
excess pressure $\Delta p$, both having their maximum (Fig. \ref{fig9}) at $n_{s}(T)\approx 1.5$ MPM for $T=-70^{0}$C. The spinodal curve
$n_{s}(T)$ of the corresponding 'liquid-gas' transition is determined by the condition $\partial \Delta \mu /
\partial n=0$. Along the curve $n_{s}(T)$, the excess compressibility $\kappa$ diverges, indicating the 'liquid-gas' phase separation for larger concentration.
The influence of the electrostatic interactions to the curve $n_{s}(T)$ is minor, they only slightly shift it towards lower concentrations. This
is consistent with our basic hypothesis: the van der Waals contribution plays the most important role in diluted MAS. The spinodal curve $n_{s}(T)$
is represented in Fig. \ref{fig4}, where we have let the van der Waals contribution independent on temperature. That gives the locus of the
thermodynamical instability of the nonmetallic phase.

At the same time, we may estimate the critical temperature $T_c^{\ell g}$ corresponding to the liquid-gas transition with the use of the
relations $\partial \Delta \mu_{tot}(T_c^{\ell g})/{\partial n} =\partial^2 \Delta \mu_{tot}(T_c^{\ell g})/{\partial n^2} = 0$. As a result, we
find $T_c^{\ell g}\sim 500^0$K and $n_c^{\ell g}\sim 11.4$ MPM. In fact, the liquid-gas separation is hidden by the quantum MIT at
$n_{c1}>n_{s}(T)$, and the actual state in competition with the low density state of localized electrons is not a dense state of solvated
electrons, but a metallic state of  free electrons.  Leaving aside details, we have estimated the critical point of the phase diagram as the
crossing between $n_{s}(T)$ and $n_{c1}(T)$. It yields $n_c(T_c) \approx 2.8$ MPM and $T_c \approx 260$ K (see Fig. \ref{fig4}). Of course, the
critical concentration $n_{s}(T)$ significatively depends on the size of ions, because the
actual packing factor increases with ion size (Fig. \ref{fig11}), whereas the critical concentration $%
n_{c}(T_{c})$ varies insignificantly with the changes in the ion size.

Using analytical expressions (\ref{disp}) and (25), we have evaluated the variation of the equilibrium cavity size (see Fig. \ref{fig10}). At
low concentrations the obtained value is close to  $r_{e}=3.2$ \AA. The cavity size decreases initially and then pronouncedly rises for metal
concentrations above $0.5$ MPM, the variation in the radius being about 5\% at $n\approx 6$ MPM. The excess molar volume, which was
experimentally measured \cite{thompson,exp}, is given by $v_{e}=4\pi (\sigma _{i}/2+r_{e})^{3}$ in our model and the concentration variations of
the cavity radius $r_{e}(n)$ essentially correspond to the the variation of the excess molar volume $v_{e}(n)$. It is interesting to notice that
the obtained  behavior of $v_{e}(n)$ is experimentally observed for excess molar volumes at temperature above $T_{c}$ \cite{thompson,exp}. We
believe that the initial decrease in $r_{e}$ may also result from formation of eventual multi-electron cavities or due to presence of metallic
droplets.

Finally, we depict the experimental phase diagram of Na-NH$_{3}$ solutions and our critical lines $n_{c1}(T)$ and $n_{s}(T)$ in Fig.
\ref{fig12}. As it is seen these lines give upper estimates with respect to the experimental ones, both for the corresponding line of
metallization and for the phase separation density range. Our calculations indicate that critical concentration $n_{s}(T)$ substantially depends
on the ion size (Fig. \ref{fig11}), the phase separation range decreases as the size of ion increases due to an enhanced excluded volume effect,
while the phase separation disappears in the case of Cs$^{+}$, which is experimentally observed \cite{thompson}.

\section{Discussion and conclusions\label{sec4}}

Using simple calculations based on a semi-continuum treatment of metal-ammonia solutions and applying the RPM for interacting excess electrons
and ions, we have investigated the insulating phase of MAS. Our study has shown that at finite metal concentrations the physics of electron
solvation is quite different from that of the infinitely diluted case. The latter is mainly determined by the cavity formation where the electron is
localized, and the excess electron can be considered from this point of view as a charge in a cavity surrounded by the oriented solvent
molecules, while the collective behavior of the electrons is mainly controlled by electrostatic interactions. However at intermediate
concentrations about several MPM, the excess electrons are governed by dispersion interactions leading to strong anomalies in the dielectric
response and concentration changes in the absorption maximum.

Another consequence of dispersion attractions between the electrons, that we have not discussed in the paper, is their plausible association.
The detailed study of such an association and related clustering effects is beyond the scope of our current work, because it requires an account
of the presence of the metallic phase. Nevertheless, our preliminary evaluations indicate that a microemulsion phase could arise in the range of
concentration $2$ MPM$<n<10$ MPM due to the small difference between the free energies of metallic and nonmetallic states. This microemulsion
phase would be characterized by a large variety of aggregates including various stripes, bubbles and so on, as reported in
numerical simulations \cite{K4}. Although magnetic properties of diluted MAS are also beyond the scope of our work, we note that the
spin-pairing of the localized electrons in the cavities could be a first stage of the association indicated above. Does the spin-paring effect
result in pure bipolaron formation or in more complicated forms? There is no clear experimental evidence. Some experimental data on the
spin-pairing were treated as an equilibrium between single electrons and spin-pairing bipolarons \cite{ccc1}, while the concentration changes in
magnetic susceptibility and excess molar volume were fitted with the use of an assumption of the formation of $e^{-}M^{+}e^{-}$ complexes
\cite{exp}. Quite generally, some deviations of our results with respect to the experimental data (on the absorption spectrum and the excess
molar volume) may be explained in terms of these possible associations. This effect complicates detailed calculations, however the formation of
such associations does not change our main conclusion about the role of induced dipole-dipole interactions, especially for the MNM transition
mechanism.

Our calculations of the dielectric, optical, and thermodynamic properties are confirmed by the experimental data, but the most remarkable
observation of the dominance of dispersion interactions is the existence of two kinds of instabilities in the nonmetallic phase. The first one
is a purely quantum transition and related with the onset of metallization, which is revealed by a sharp increase in the static dielectric
constant when approaching the polarization catastrophe density. Although we have considered here only the nonmetallic phase, we think that the
theory of the MNM transition in MAS is actually more complicated and provided by a combination of exchange-correlations effects in the metallic
phase and dipolar interactions between solvated electrons in the insulating phase. This will be the scope of our future papers. The second
transition is classical and is revealed by a phase separation in MAS at low temperatures. This instability is similar to that observed in
classical fluids with van der Waals attractions. These attractions become dominant at large enough concentrations. At the same time, in our case
of MAS the metallic phase is also instable at low temperatures for light alkali metals as we have shown in \cite{comptes}. Therefore, the phase
separation in MAS has a dual nature. An additional argument in favor of our dual scenario is that both instabilities are to disappear in
$Cs-NH_{3}$ solutions. The latter is experimentally confirmed by the absence of the phase separation in these solutions.

The dominance of dispersion attractions also results in the negativity of dielectric function $\varepsilon (\omega )$ over a wide frequency range
at $k=0$. At this stage, we have not extended our calculations at $k \ne 0$, but $\epsilon(0, \bf{k} \ne 0)$ should be also negative
\cite{quem1,quem2,quem3}. Together with the instability of the metallic phase, that may lead to unusual properties of MAS. For example,
frustrating the phase separation in fast frozen metal-ammonia solutions, as did Ogg \cite{s1} a long time ago, the negativity of the dielectric
function may result in unusual electronic states with strong attractions \cite{quem2}. Is it superconducting or highly conducting but
nonstationary the answer depends on dynamics and conditions of the electron localization, and the existing experimental data on anomalous
conductivity in frozen MAS has been debated during decades \cite {s1,s2,s3}. However, this question of possible negativity of the dielectric
constant has been recently revived by its experimental observation in an expanding metal (rubidium) \cite{matsuda1}, as well as by the
observation of a first-order MNM transition in expanded fluid Hg \cite{matsuda2}.

Finally, we think that the proposed scenario may be applied to systems involving
interacting quantum particles having a strong polarizability. In our model the
behavior of quantum polarizable particles is controlled by two dimensionless
parameters: $c_{1}=\alpha _{0}/r_{e}^{3}\varepsilon _{\infty }$ and $%
c_{2}=8nr_{e}^{3}$. The first of them is the ratio between the polarizability to the excluded volume, while the second one is the relative
fraction of the excluded volume in the liquid. The parameter $c_{1}$ is small in the case of ordinary polarizable fluids, hence the parameter
$c_{2}$ should be close to unity to provoke the polarization catastrophe and the present scenario remains unlikely for such systems. However in
case of self-trapped quantum particles like excitons and polarons,  $c_{1}$ may become large enough to provoke a polarization catastrophe at low
concentrations opening a region in which both the insulating and the metallic phases are not stable. The metallization may be associated with a
phase separation in that case. Hence, our scenario is quite general and may take place in other systems, for instance, in alkali metal-alkali
halide solutions where solvated electrons, phase separation, and dielectric anomalies were experimentally observed \cite{Freyland}. This dual
nature could be also the origin of the existence of two critical points in the phase diagram of excitons in semiconductors \cite{scho,smith}.

At this stage the model is not complete, since we have restricted ourselves to low metal concentrations. Nevertheless, our calculations have
indicated that MAS is an example of quantum-classical system whose thermodynamic and dielectric properties are controlled by dispersion
interactions between self-trapped quantum quasiparticles. We believe that our scenario is more general and may be revealed for other quantum
particles such as excitons in semiconductors and polarons in oxides, which may provide a new insight into metal-insulator transitions in
condensed matter physics.

\begin{acknowledgements}
G.N.Ch. thanks the Leverhulme Trust and Russian Foundation for Basic  research for partial support of this work.
\end{acknowledgements}

\section*{Appendix 1. Free energy of excess electrons in the insulting phase}

We treat MAS as a system consisting of $N$ excess electrons, $N$ of monovalent metal ions, and $N_{s}$ classical solvent particles, whose
distribution depends not only on their coordinates $\mathbf{R}^{\{N_{s}\}}=\{%
\mathbf{R}_{1},\mathbf{R}_{2},....\mathbf{R}_{N_{s}}\mathbf{\}}$ but also on
the orientations $\mathbf{w}^{\{N_{s}\}}=\{\mathbf{w}_{1},\mathbf{w}_{2},....%
\mathbf{w}_{N_{s}}\mathbf{\}}$ of their dipole momenta $\mathbf{m}$. The system of interacting classical particles and electrons is described by
the grand partition function $\Xi $ given by
\begin{equation}
\Xi =\left\langle \left\langle \exp [-\beta (H-\mu _{e}N-\mu _{i}N-\mu _{s}N_{s})]\right\rangle _{s}\right\rangle _{e+i},  \label{2a}
\end{equation}
where the symbols $\left\langle ...\right\rangle _{s}$and $\left\langle ...\right\rangle _{e+i}$ denote the averages over solvent, electronic,
and ionic degrees of freedom respectively, $H$ is the total Hamiltonian of the system, $\mu _{e}$, $\mu _{i}$, and $\mu $are the chemical
potentials of electrons, ions, and solvent particles. We write the total Hamiltonian as the sum of electronic ($H_{e}$), ionic ($H_{i}$) and
solvent ($H_{s}$) contributions:
\begin{eqnarray}
&&H =T+\sum_{ij}^{NN_{s}}u_{es}(\mathbf{r}_{-i}\mathbf{-R}_{sj},\mathbf{w}%
_{j})+\frac{1}{2\varepsilon _{\infty }}\sum_{i\neq j}^{N}\frac{1}{|\mathbf{r}%
_{-i}\mathbf{-r}_{-j}|}+\sum_{i\neq j}^{N}u_{ei}(\mathbf{r}_{-i}\mathbf{-R}%
_{+j})+  \label{11}\\
&&\frac{1}{2}\sum_{i\neq j}^{N}u_{ii}(|\mathbf{R}_{+i}\mathbf{-R}%
_{+j}|)+\sum_{ij}^{NN_{s}}u_{is}(\mathbf{R}_{+i}\mathbf{-R}_{sj},\mathbf{w}%
_{j})+\frac{1}{2}\sum_{i\neq j}^{N_{s}}u_{ss}(\mathbf{R}_{si}\mathbf{-R}%
_{sj},\mathbf{w}_{i}-\mathbf{w}_{j}), \nonumber
\end{eqnarray}
The first term in the right side of (\ref{11}) is the kinetic energy of electrons, the second one is due to electron-solvent interactions, the
next terms are the electrostatic interactions between electrons and ions, while the last two terms result from the ion-solvent and
solvent-solvent contributions, respectively. The electron-solvent term includes electrostatic as well as nonelectrostatic contributions. The
similar contributions are in the ion-ion ($u_{ii}$), the ion-solvent ($u_{is}$), and
the solvent-solvent ($u_{ss}$) potentials. We take into account in (%
\ref{11}) that the interactions between electrons are screened by the high-frequency dielectric constant $\varepsilon _{\infty}$.

In principle, the average over quantum and classical degrees of freedom can
be provided with the use of $N-$electron wave function $\Psi (\mathbf{r}_{1},%
\mathbf{r}_{2}...\mathbf{r}_{N})$ or expressed in terms of path integrals, however neither the computation of the electron wave functions nor
the evaluation of path integrals is easy to perform. The most popular version for treating electron-electron interactions is based on the local
density approximation \cite{Parr} in which the interactions are represented by the sum of electrostatic interactions between classical charges
related with electron density $n_{e}(\mathbf{r})$and a short-range contribution due to change in the chemical potential $\mu
_{e}(n_{e}(\mathbf{r}))$ caused by exchange-correlation effects. Unfortunately, such way is not suitable for our case, because it ignores the
long-range nature of electron-electron correlations. Instead of it, we take into that the electrons are localized and their deviations
$\mathbf{u}_{j}$ from their equilibrium positions $\mathbf{R}_{-j}=\mathbf{r}_{-j}-\mathbf{u}_{j}$ of their centers of mass are small, i.e.,
$\mathbf{u}_{j}<<\mathbf{R}_{-j}$. Then expanding the electron-electron and electron-ion interactions with respect to this small parameter and
restrict ourselves by the dipole approximation, we obtain
\begin{eqnarray}
\sum_{i\neq j}^{N}\frac{1}{|\mathbf{r}_{-i}\mathbf{-r}_{-j}|} &\approx
&\sum_{i\neq j}^{N}\{\frac{1}{|\mathbf{R}_{-i}\mathbf{-R}_{-j}|}+\frac{2%
\mathbf{u}_{j}\cdot \nabla }{|\mathbf{R}_{-i}\mathbf{-R}_{-j}|}+\mathbf{u}%
_{j}\cdot \nabla \frac{2}{|\mathbf{R}_{-i}\mathbf{-R}_{-j}|}\nabla \cdot
\mathbf{u}_{i}\},  \label{dipole} \\
\sum_{i\neq j}^{N}\frac{1}{|\mathbf{R}_{+i}\mathbf{-r}_{-j}|} &\approx
&\sum_{i\neq j}^{N}\{\frac{1}{|\mathbf{R}_{+i}\mathbf{-R}_{-j}|}+\frac{%
\mathbf{u}_{j}\cdot \nabla }{|\mathbf{R}_{+i}\mathbf{-R}_{-j}|}+\mathbf{u}%
_{j}\cdot \nabla \frac{1}{2|\mathbf{R}_{+i}\mathbf{-R}_{-j}|}\nabla \cdot \mathbf{u}_{j}\}.  \label{ch-dip}
\end{eqnarray}
The first terms in these relations are the Coulomb interactions between classical charges (cations and anions), the second terms are responsible
for interactions between permanent dipoles and classical charges, while the last term in (\ref{dipole}) is due to interactions between induced
dipoles, but the last term in (\ref{ch-dip}) is interactions between classical charges and induced dipoles. The contribution responsible for
interactions between the permanent dipoles and classical charges vanishes after averaging over quantum degrees of freedom, because the localized
electrons have spherical symmetry in the ground state and their average momenta $\left\langle \mathbf{m}\right\rangle \propto \left\langle
\mathbf{u}\right\rangle $ are equal to zero. The
similar situation for the contribution caused by the last term in (\ref{ch-dip}%
),  it disappears after averaging over the classical degrees of freedom due to isotropic distribution of classical charges. As a result,
expansions (\ref {dipole}) and (\ref{ch-dip})) provide only the classical electrostatic interactions (the first terms in (\ref
{dipole})-(\ref{ch-dip})) and the induced dipolar interactions (the last term in (\ref{dipole})).

The set $\{\mathbf{R}_{-}^{\{N\}},\mathbf{R}_{+}^{\{N\}},\mathbf{R}%
_{s}^{\{N_{s}\}},\mathbf{w}_{i}^{\{N_{s}\}}\}$ corresponds to classical degrees of freedom. Averaging over the classical degrees, we obtain an
effective hamiltonian $H_{ef}$ depending on correlation functions between solvent and excess charges and on quantum coordinates. The evaluation
of the correlation functions can be performed by the theory of integral equations of classical liquids \cite{Hans}. Such approach has been
applied in \cite{25,c4} to model a mixture of dipolar solvent particles, excess electrons, and ions. However, these equations can be solved only
numerically, and to simplify the calculations, we apply the RPM and treat the solvation of excess charges as a formation of cavities and
screening of
charges localized in the cavities. In this case the effective hamiltonian $%
H_{ef}$ is written as
\begin{equation}
H_{ef}=Nh_{0}+\Delta F_{cl}+\frac{1}{\varepsilon _{\infty }}\sum_{i\neq j}^{N}\mathbf{u}_{i}\cdot
\mathbf{T}(\mathbf{R}_{-i}-\mathbf{R}_{-j})\cdot \mathbf{u}_{j},  \label{ef1}
\end{equation}
where $h_{0}$ is the one-electron contribution for noninteracting electrons, $\Delta F_{cl}$ is the change in the free energy due to the
classical interactions between solvated charges (cations and anions), while the last term corresponds to dispersion interactions between the
electrons.

The average of the one-electron hamiltonian $h_{0}$ over quantum degrees of freedom can be obtained by the integral equation methods for an
excess electron in a polar liquid \cite{chuev,cukier}, and the free energy $f_{0}$ can be expressed \cite{Bip} as:
\begin{equation}
f_{0}=-\frac{1}{2}\int \Psi _{0}(\mathbf{r})\nabla ^{2}\Psi _{0}(\mathbf{r})d%
\mathbf{r}+\frac{1}{2}\sum_{ij}\int |\Psi _{0}^{2}(\mathbf{r})|u_{ef}(%
\mathbf{r-r}_{1})|\Psi _{0}^{2}(\mathbf{r}_{1})|d\mathbf{r}d\mathbf{r}%
_{1}+f_{cav},  \label{6}
\end{equation}
where $\Psi _{0}(\mathbf{r})$ is the electron wave function in the ground state, $f_{cav}$ is the nonelectrostatic contribution due to cavity
formation, while $u_{ef}(\mathbf{r})$ is the effective interaction potential due to polarization of solvent molecules. It can be found
\cite{UFN} that
the long-range part of the effective potential has the asymptotic $%
u_{ef}(r\rightarrow \infty )=-(\varepsilon _{\infty }^{-1}-\varepsilon _{s}^{-1})/r$, while cavity contribution can be parameterized as a sum of
volume and surface terms \cite{Bip}. Finally, using the semi-continuum
approximation for the effective potential $u_{ef}(\mathbf{r})$, we obtain (%
\ref{1}), while the change $\Delta F/N$ in the total free energy is written by (\ref{total}).

\section*{Appendix 2. The effective polarizability of a quantum polarizable fluid}

Let us consider a large sample of $M>>1$ Drude oscillators in a volume $V$ with surface $S$ and extract a sphere containing $N$ particles  from
this large sample (with $M>>N>>1$). In the spirit of the Clausius-Mossotti calculation, we may evaluate the polarization of the independent
sphere subjected to an external field, which is nothing but the Lorentz local field (\ref{localF}). The momentum induced by the external field
on $i$-th particle is:
\begin{equation}
\mathbf{m}_{i}(\omega \mathbf{)}=\sum_{j}^{N}[\alpha _{0}^{-1}\mathbf{I-}%
\varepsilon _{\infty }^{-1}\mathbf{T}_{i}{}_{j}\mathbf{]}^{-1}\cdot \mathbf{%
E}_{\ell}(\omega),
\end{equation}
while the expression for the average moment $\left\langle \mathbf{m}(\mathbf{%
\omega )}\right\rangle $ is written as
\begin{equation}
\left\langle \mathbf{m}(\omega \mathbf{)}\right\rangle =\left\langle
\sum_{j}^{N}[\alpha _{0}^{-1}\mathbf{I-}\varepsilon _{\infty }^{-1}\mathbf{T}%
_{ij}\mathbf{]}^{-1}\right\rangle \cdot \mathbf{%
E}_{\ell}(\omega)=\alpha
(\omega )\mathbf{%
E}_{\ell}(\omega)/\varepsilon _{\infty },
\end{equation}
where $\left\langle \mathbf{...}\right\rangle $ is the average over distributions of particles, while the last identity in the above relation is
the definition of renormalized effective polarizability $\alpha (\omega )$ in which we account that the external field is screened by the
solvent containing in the sample. Then the
change $\Delta \mathbf{P}$ in the macroscopic polarization, determined as $%
\Delta \mathbf{P=}\sum_{i}^{N}\mathbf{m}_{i}(\omega \mathbf{)}$, is written in terms of effective polarizability by (\ref{Pdip2}), that allows
us to obtain (\ref{Cl-Mo}).

The averaging procedure results in the self-consistent equation for the renormalized polarizability written formally as (\ref{effectivealpha}).
It can be evaluated by the integral equations method \cite {hoye,wertheim}, while the Pad\'{e} approximation \cite {pratt,chandler1,shweizer}
provides the analytical expression (\ref{pade}) for this function. Substituting it into (\ref{effectivealpha}), we obtain quadratic equation
(\ref{quadra}) for $\alpha (\omega )$ whose solution is complex, while their real and imaginary parts are given by:
\begin{equation}
\alpha _{r}(\omega )=\alpha _{0}B(\omega )A^{-1}(\omega )[\sqrt{1-2A(\omega
)B^{-2}(\omega )}-1],
\end{equation}
\begin{equation}
\alpha _{i}(\omega _{-}<\omega <\omega _{+})=\alpha _{0}A^{-1}(\omega )\sqrt{%
2A(\omega )-B^{2}(\omega )},  \label{ri}
\end{equation}
where $B(\omega )=\omega ^{2}\omega _{0}^{-2}-1+b\alpha _{0}\varepsilon
_{\infty }^{-1}$, $A(\omega )=2\alpha _{0}\varepsilon _{\infty
}^{-1}b[\omega ^{2}\omega _{0}^{-2}-1+ab^{-1}\alpha _{0}\varepsilon _{\infty
}^{-1}]$, and the functions $a$ and $b$ are determined in terms of
dimensionless parameter $x=8nr_{e}^{3}$ as:
\begin{equation}
a(x)=\frac{\pi x}{24r_{e}^{6}}[\frac{1-0.3168x-0.3205x^{2}+0.1078x^{3}}{%
(1-0.5236x)^{2}}]=\frac{\pi n}{3r_{e}^{3}}I_{0}(x),
\end{equation}
\begin{equation}
b(x)=0.72505n\frac{(2.70797+1.68918x-0.3157x^{2})}{%
(1-0.59056x+0.20059x^{3})I_{0}(x)}=\frac{5\pi nI_{1}(x)}{8I_{0}(x)}.
\end{equation}
The root of (\ref{ri}) is equal zero at edge eigen-frequencies $\omega _{-}$%
and $\omega _{+}:$
\begin{equation}
\omega _{\pm }^{2}(n)=\omega _{0}^{2}+b(n)\varepsilon _{\infty }^{-1}\pm
2a^{1/2}(n)\varepsilon _{\infty }^{-1},
\end{equation}
determining the low and the high edge frequencies of the DOS spectrum.
Hence, we can calculate effective polarizability $\alpha (\omega )$ and the DOS $D(\omega)$ for a given concentration $n$
and frequency $\omega _{0}$, while the later can be derived from $r_{e}$ by (%
\ref{omega}).

The solution of (\ref{quadra}) depends on the two dimensionless parameters $%
c_{1}=1/\varepsilon _{\infty }r_{e}^{3}\omega _{0}^{2}$ and $%
c_{2}=8nr_{e}^{3} $. To obtain analytical estimations of the expressions, we consider the limiting case $c_2 \rightarrow 0$, or more precisely
$r_e \rightarrow 0$ (called as the point oscillator limit). We have $a=8\pi n\int_{2r_{e}}^{\infty }r^{-4}dr=\pi
n/3r_{e}^{3}$ and $b=5\pi n/8$, which leads to expressions (\ref{alil}) and (%
\ref{edgl}) for the imaginary part of the effective polarizability and the edge frequencies. At the same time, the real part $\alpha _{r}(0) $
becomes in the same limit:
\begin{equation}
\alpha _{r}(0,n\rightarrow 0)=\alpha _{0}[1+\frac{\pi n\alpha _{0}^{2}(0)}{%
3\varepsilon _{\infty }^{2}r_{e}^{3}}],  \nonumber
\end{equation}
while it is equal to $\alpha _{r}(0,n_{c1})=2\alpha _{0}/(1-5\pi
n_{c1}\alpha _{0}/8\varepsilon _{\infty })$ at the critical concentration $%
n_{c1} $. The instability occurring at $n_{c1}$ is given by $\omega_-(n_{c1})=0$, which leads to:
\begin{equation}
n_{c1}=\frac{3\varepsilon _{\infty }^{2}r_{e}^{3}\omega _{0}^{4}}{4\pi }=%
\frac{C_{1}(T)}{(2r_{e})^{3}},
\end{equation}
where $C_{1}(T)=(6/\pi )(\varepsilon _{\infty }C_{r}^{2}(T)/r_{e})^{2}$ is the numerical factor weakly depending on temperature. This factor
$C_{1}$ is about of $1/3$ at low temperatures, and hence the critical concentration is roughly $n_{c1}\approx 1/3(2r_{e})^{3}$. Substituting the
analytical expressions for $n_{c1}$ and $\alpha _{r}(0,n_{c1})$ we may evaluate the second instability occurring at $n_{\infty}$ given by:
\begin{equation}
\frac{n_{\infty }}{n_{c1}}=\frac{\alpha _{0}(1-5\pi n_{c\infty }\alpha _{0}/8\varepsilon _{\infty })}{2\varepsilon _{\infty }r_{e}^{3}}.
\end{equation}
The two instabilities in this limit are found to be quite close to that of metallization, since $n_{\infty }/n_{c1}\approx 0.92$ at low
temperatures.

Now we evaluate the zero-point energy of the collective mode in the limiting case. Substituting
 (\ref{alil}) and (\ref{edgl}) into the expression of the
zero-point vibrations, we have
\begin{equation}
\Delta f_{d}=\frac{9\varepsilon _{\infty }^{2}r_{e}^{3}}{2\pi ^{2}n}%
\int_{\omega _{-}}^{\omega _{+}}\omega ^{2}(\omega _{+}^{2}-\omega ^{2})^{1/2}(\omega ^{2}-\omega _{-}^{2})^{1/2}d\omega -\frac{3\omega
_{0}}{2}.
\end{equation}
Introducing the new variable $y$, which is related with the frequency $%
\omega $ as
\begin{equation}
\omega ^{2}=\omega _{0}^{2}+\delta \cos y,
\end{equation}
where $\delta =(4\pi n/3\varepsilon _{\infty }^{2}r_{e}^{3})^{1/2}$, we can rewrite the integral as
\begin{equation}
\Delta f_{d}=\frac{3\omega _{0}}{\pi }\int_{0}^{\pi }[(1+\delta \omega _{0}^{-2}\cos y)^{1/2}-1]\sin ^{2}ydy.
\end{equation}
Expanding the root in the integrand into the series with respect to
parameter $\delta \omega _{0}^{-2}$, i.e. $(1+\delta \omega _{0}^{-2}\cos
y)^{1/2}-1=\delta \omega _{0}^{-2}\cos y/2-(\delta \omega _{0}^{-2}\cos
y)^{2}/8+...$, we obtain
\begin{equation}
\Delta f_{d}=-\frac{3\omega _{0}}{\pi }[\frac{\pi \delta ^{2}}{64\omega _{0}^{4}}+\frac{5\pi \delta ^{4}}{64\cdot 32\omega _{0}^{8}}+O(\delta
^{6}\omega _{0}^{-12})].  \label{disp1}
\end{equation}
Expressing $\delta $ via density $n$, we have the final expression (\ref
{disp}) for the dispersion contribution. Completing the consideration of the
limiting case, we indicate that the above estimate for the energy of
zero-point vibrations is only asymptotically valid at $c_{2}\rightarrow 0$,
however we should account the higher contributions at finite $n$, moreover
the series is divergent at the point $n_{c1}$ of the polarization
catastrophe, since $\delta \omega _{0}^{-2}=1$.

\newpage
\begin{figure}[tbp]
\vspace{14cm}
\caption{Schematic sketch of model. The solvated cations with charge $+e$ are represented by the orange circles and have cartesian coordinates
$\bf{R}_{+j}$ of their centers of mass, whereas the cavities with localized electrons are represented by the blue circles and have the
coordinates $\bf{R}_{-i}$ of their centers of mass. The instantaneous position of $i-$th electron is $\bf{r}_{-i} = \bf{R}_{-i}+\bf{u}_i$, where
$\bf{u}_i$ is the relative coordinate of the electron with respect to the cavity center. Both cations and solvated electrons are assumed to have
the same size $\sigma$. The inset represents the solution in which the solvent is treated as a continuum medium with dielectric constant
$\epsilon_{NH_3}(\omega)$.} \label{fig0}
\end{figure}

\newpage
\begin{figure}[tbp]
\includegraphics{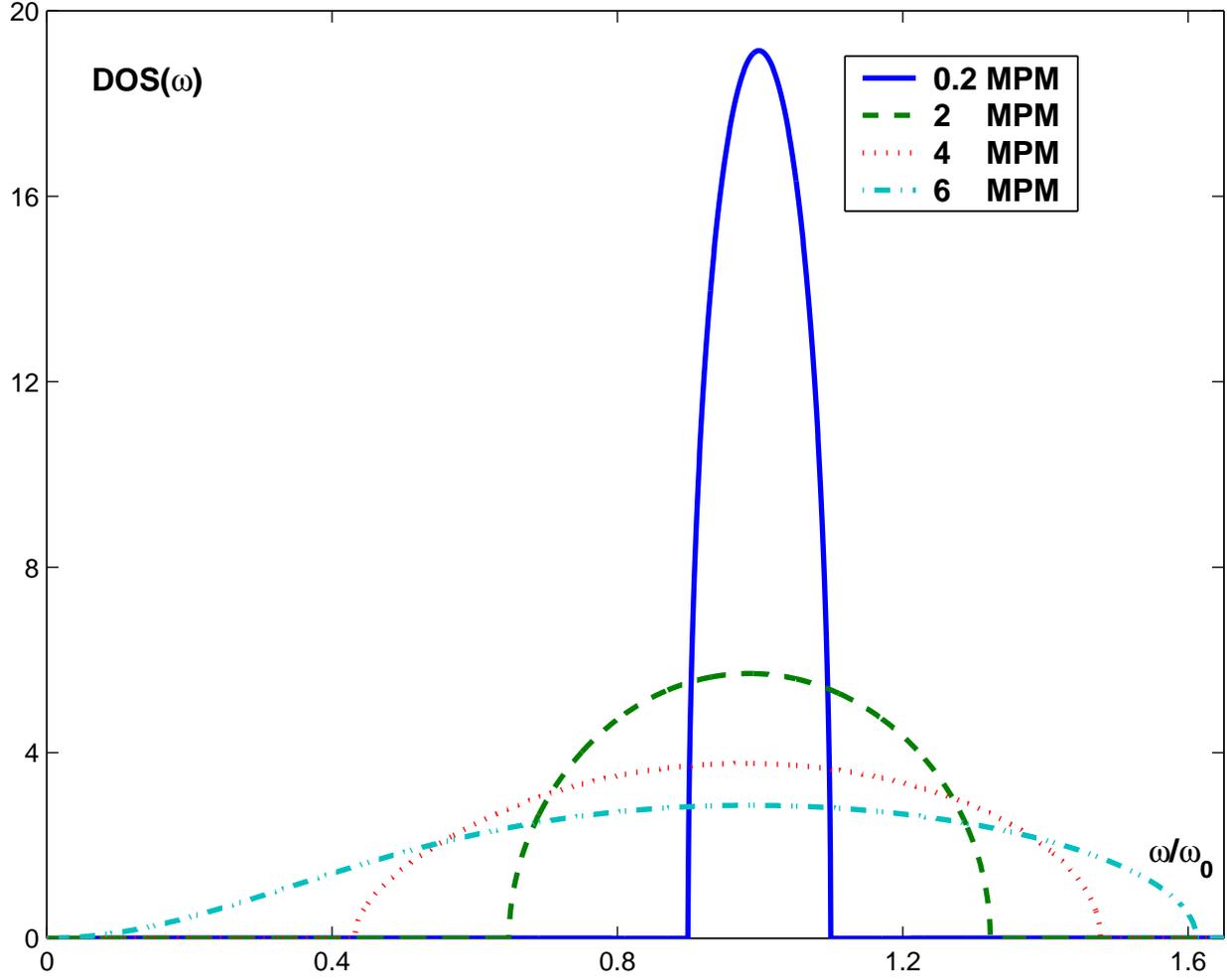}
\caption{ The dimensionless DOS $D(\protect\omega ,n)\protect\omega_{0}$ versus dimensionless frequency $\omega/\protect\omega_{0}$ at
various metal concentration $n$, $T=-70^0$C, and fixed $\protect\omega%
_{0}=0.9$ eV and $r_{e}=3.2$\AA. The values of metal concentration $n$ are indicated at the corresponding lines.} \label{fig1}
\end{figure}

\begin{figure}[tbp]
\includegraphics{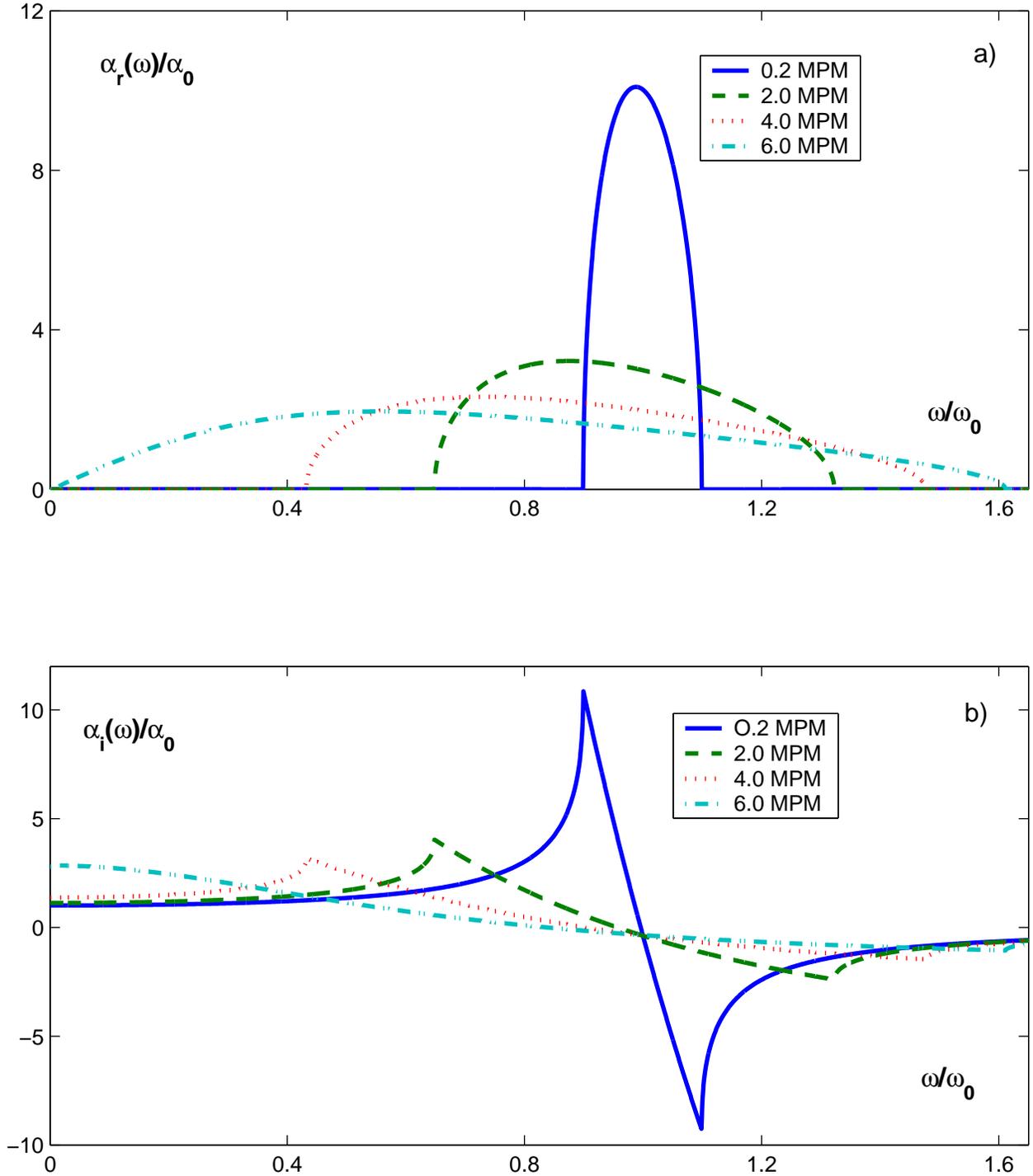}
\caption{ The dimensionless imaginary part $\alpha _{i}(\omega )/\alpha_0(0)$ (a) and the real part $\alpha _{r}(%
\protect\omega )/\protect\alpha_0(0)$ (b) of effective polarizability at
various metal concentration $n$, $T=-70^0$C, and fixed $\protect\omega%
_{0}=0.9$ eV and $r_{e}=3.2$\AA. The notions are the same as in Fig. \ref{fig1}.} \label{fig2}
\end{figure}

\begin{figure}[tbp]
\includegraphics{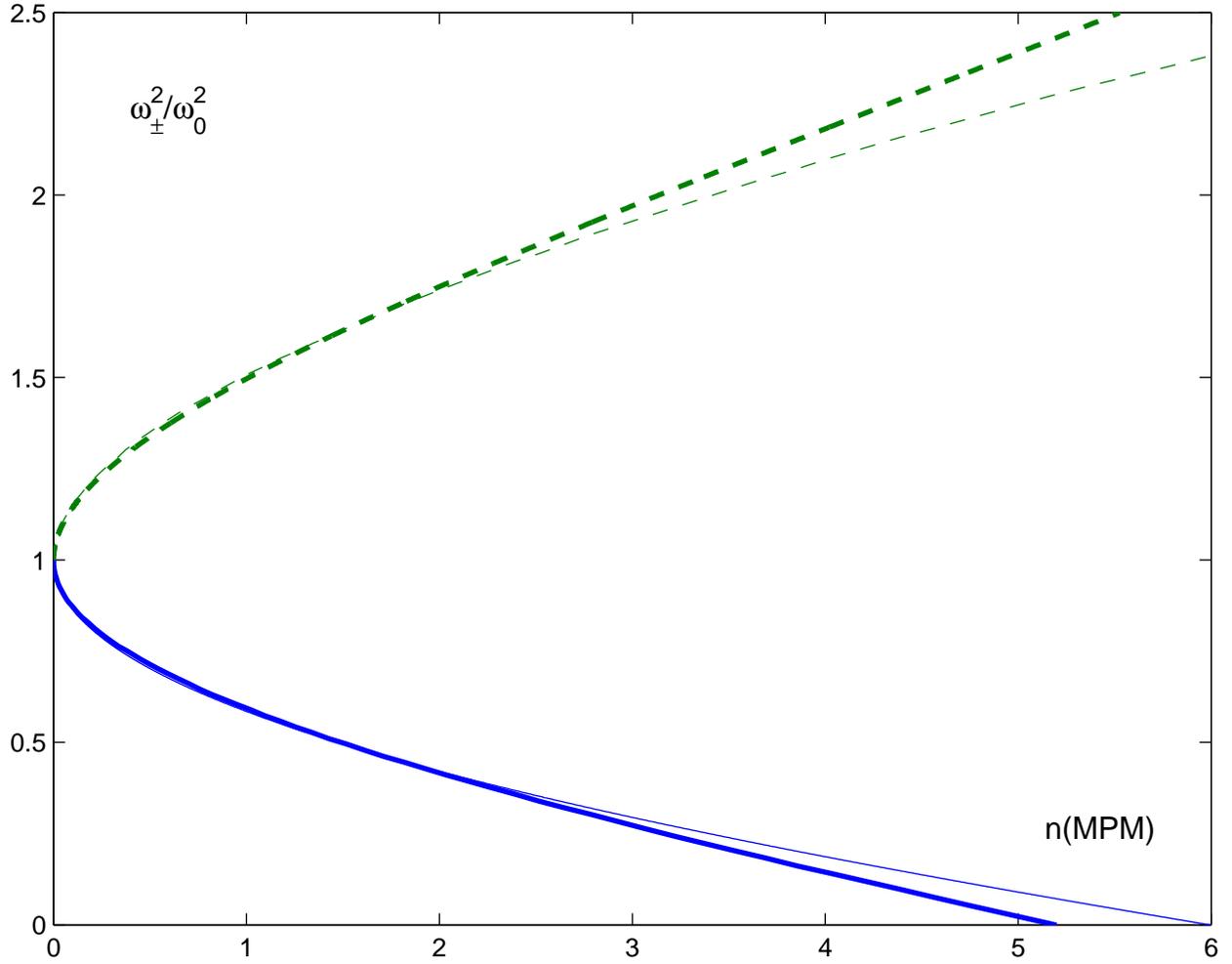}
\caption{The dimensionless edge square frequencies $\omega^2_{-}/%
\protect\omega^2_{0}$ (solid lines) and $\omega^2_{+}/\protect\omega%
^2_{0}$ (dashed lines) versus metal concentration at fixed cavity radius $%
r_{0} $ (thick lines) and with the account of concentration dependence of the radius $r_e(n)$ (thin lines). The notions are the same as in Fig.
\ref{fig1}.} \label{fig3}
\end{figure}

\begin{figure}[tbp]
\includegraphics{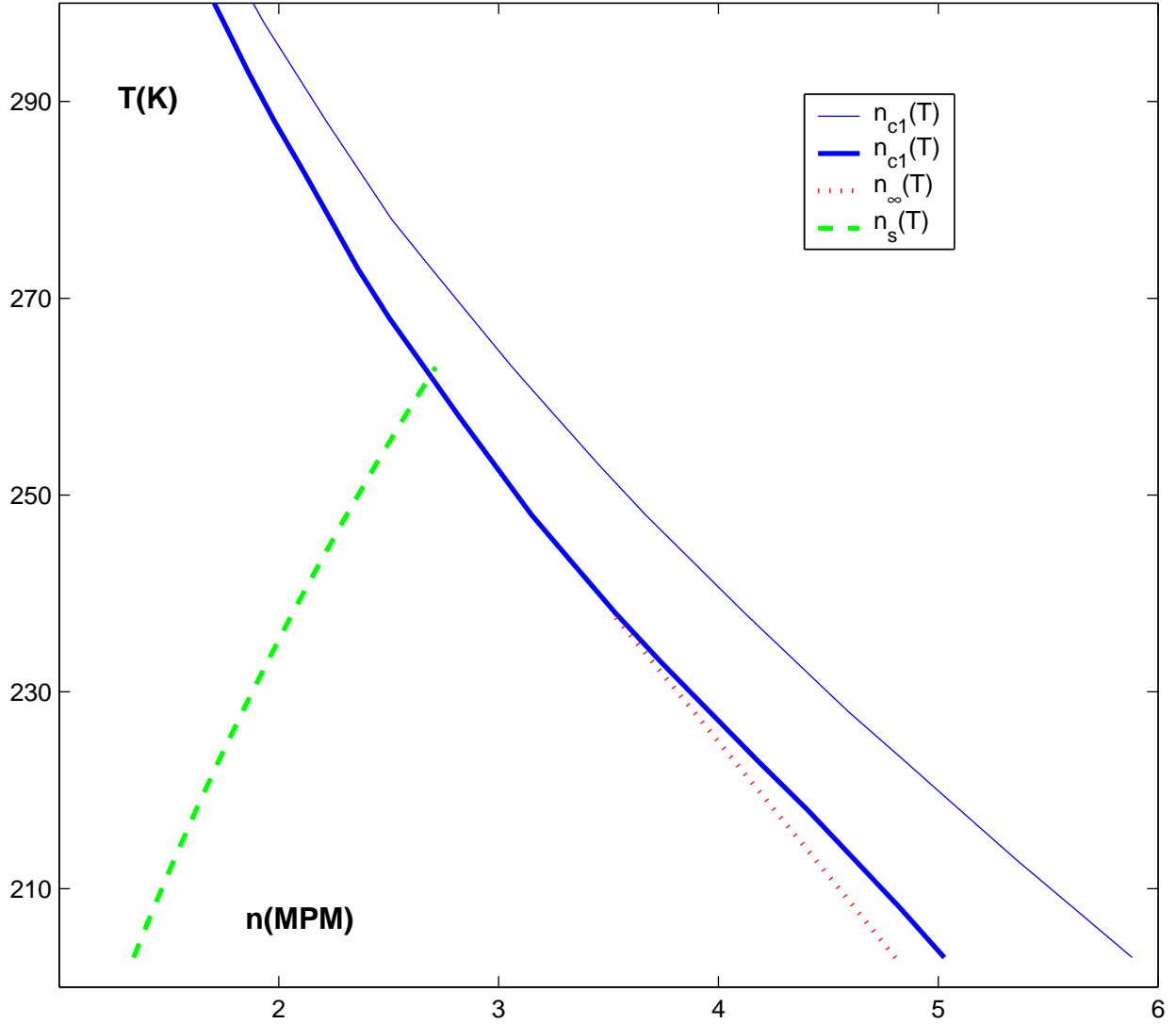}
\caption{Critical lines for Na-NH$_{3}$ solutions. The solid lines correspond to the instability concentration $n_{c1}(T)$, the dashed ones to
the spinodal line $n_s(T)$ (see text), the dotted line to divergent static dielectric constant (polarization catastrophe). The thick solid line
indicates $n_{c1}(T)$ calculated at fixed cavity radius $r_{e}=3.2$ \AA\ and frequency $\protect\omega_{0}(T)$, while the thin one to that with
the account of the concentration dependencies of the above parameters.} \label{fig4}
\end{figure}

\begin{figure}[tbp]
\includegraphics{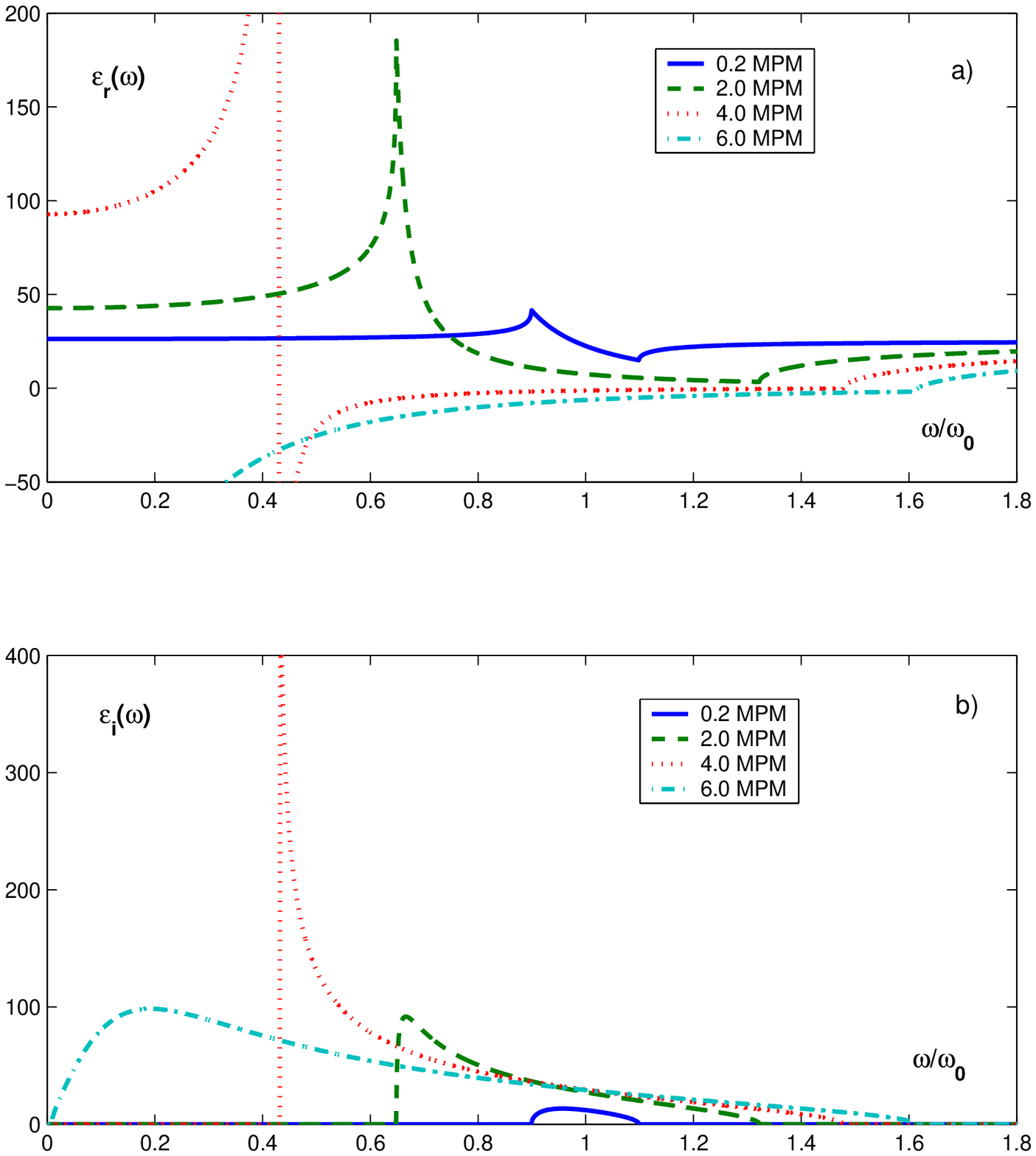}
\caption{The real part $\protect\epsilon _{r}(\protect\omega )$ (a) and the imaginary part $\protect\epsilon _{i}(\protect\omega )$ (b) of the
dielectric function at various metal concentration $n$, $T=-70^0$C, and fixed $\protect\omega_{0}=0.9$ eV and $r_{e}=3.2$\AA. The notions are
the same as in Fig. \ref{fig1}.} \label{fig5}
\end{figure}

\begin{figure}[tbp]
\includegraphics{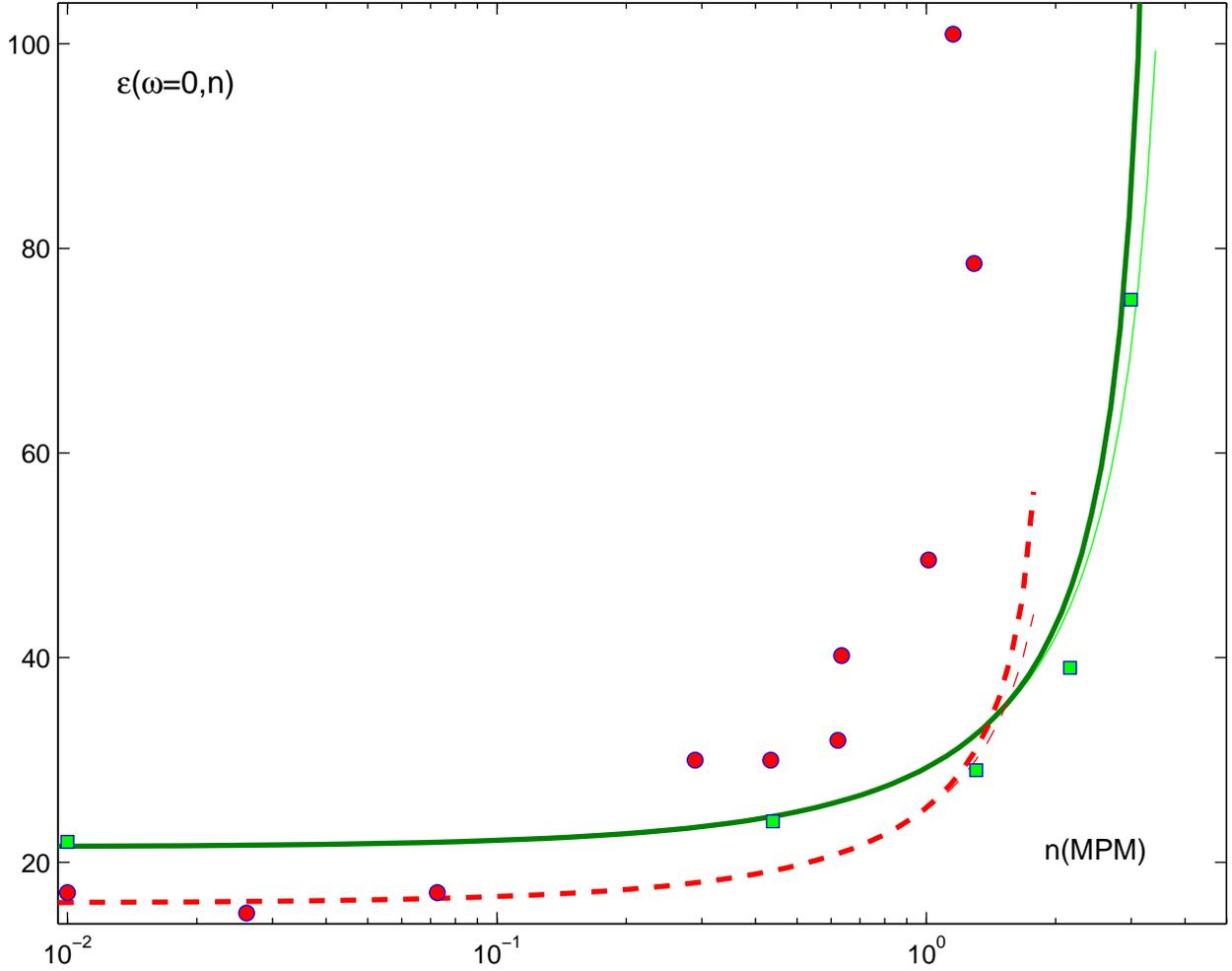}
\caption{The static dielectric constant $\protect\epsilon _{r}(\protect%
\omega =0)$ versus the metal concentration; the circles symbols correspond
to the experimental data \protect\cite{pk} on the low-frequency dielectric
constant at $T=+20^{o}$C, the squares to that at $T=-35^{o}$C \protect\cite
{pk1}. The dashed and solid curves show our results at $T=+20^{o}$C and $%
-35^{o}$C,respectively. The thin and the thick curves correspond to calculations at fixed cavity radius $r_e$ and with the account of the
concentration dependence of the radius $r_e(n)$, respectively.} \label{fig6}
\end{figure}

\begin{figure}[tbp]
\includegraphics{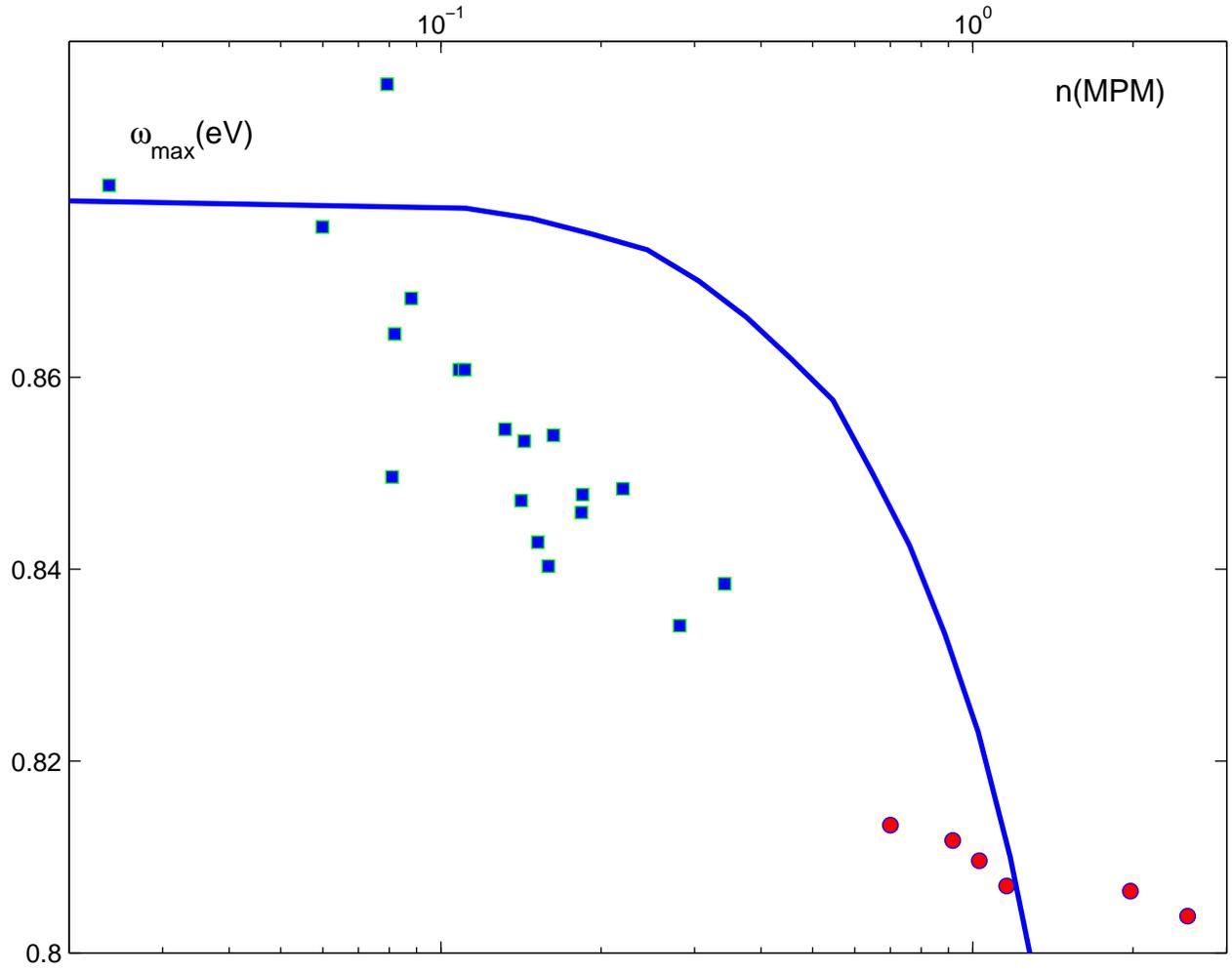}
\caption{ The locus of the maximum $\protect\omega _{\max }$ of optical absorption in MAS, the triangle symbols indicate the experimental data
on the absorption maximum in Na-NH$_{3}$ at $T=-65^{o}$C obtained from \protect\cite{ccc} and square symbols from \protect\cite{ccc1}, while the
solid curve shows our results at the same temperature. } \label{fig7}
\end{figure}

\begin{figure}[tbp]
\includegraphics{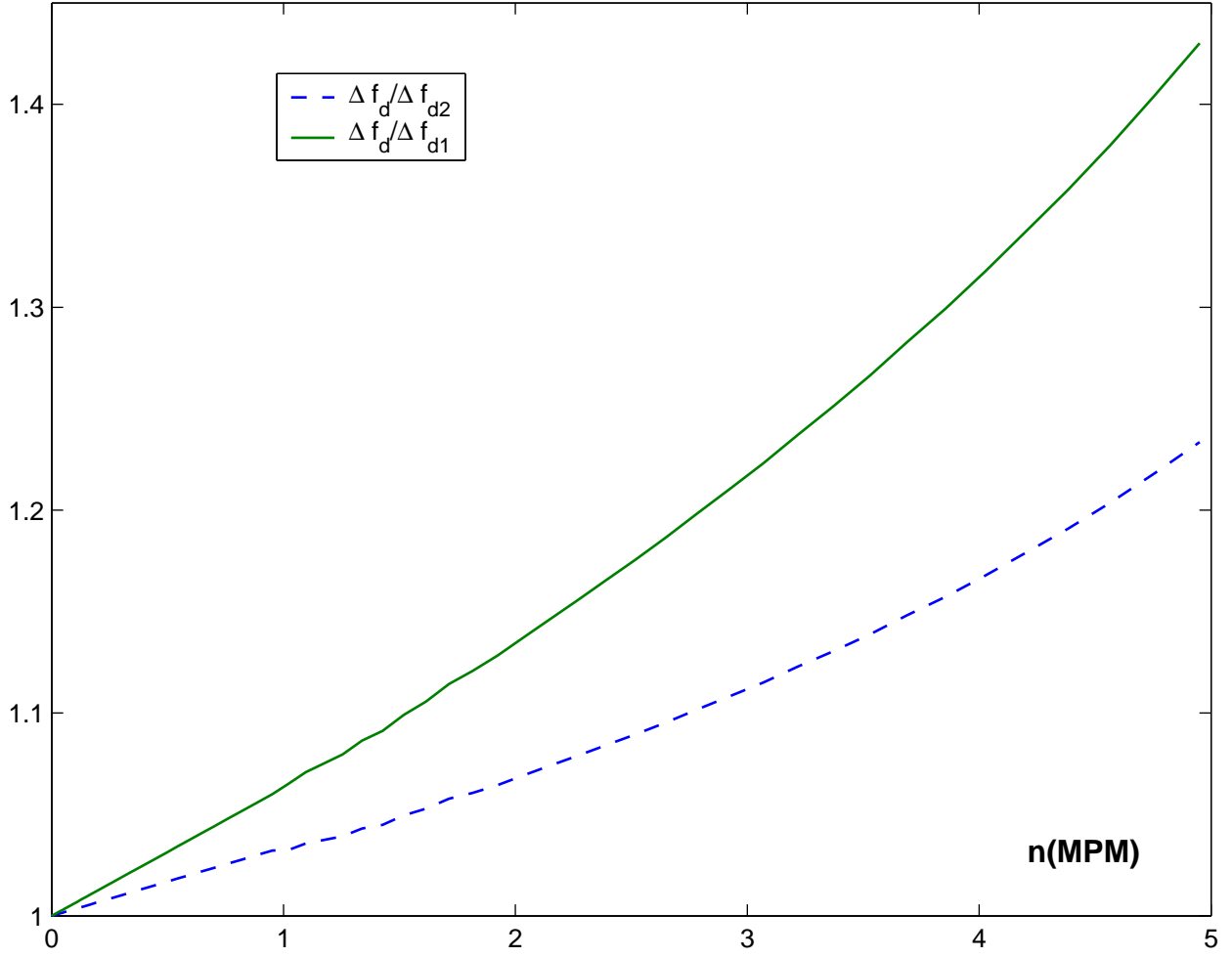}
\caption{Concentration dependence of the dispersion contribution $\Delta f_{d}(n)$ calculated at $T=-70^0$C. The dashed line corresponds to the
contribution normalized with respect to its limiting values with account the first-order correction $\Delta f_{d1}=-\pi n\omega
_{0}\alpha_{0}^{2}/16\varepsilon _{\infty }^{2}r_{e}^{3}$, while the solid line to that with respect to the second-order correction $\Delta
f_{d2}=f_{d1}(1+5\pi n\alpha_{0}^{2}/24\varepsilon _{\infty }^{2}r_{e}^{3})$. All other conditions are the same as that in Fig. \ref{fig1}.}
\label{fig8}
\end{figure}

\begin{figure}[tbp]
\includegraphics{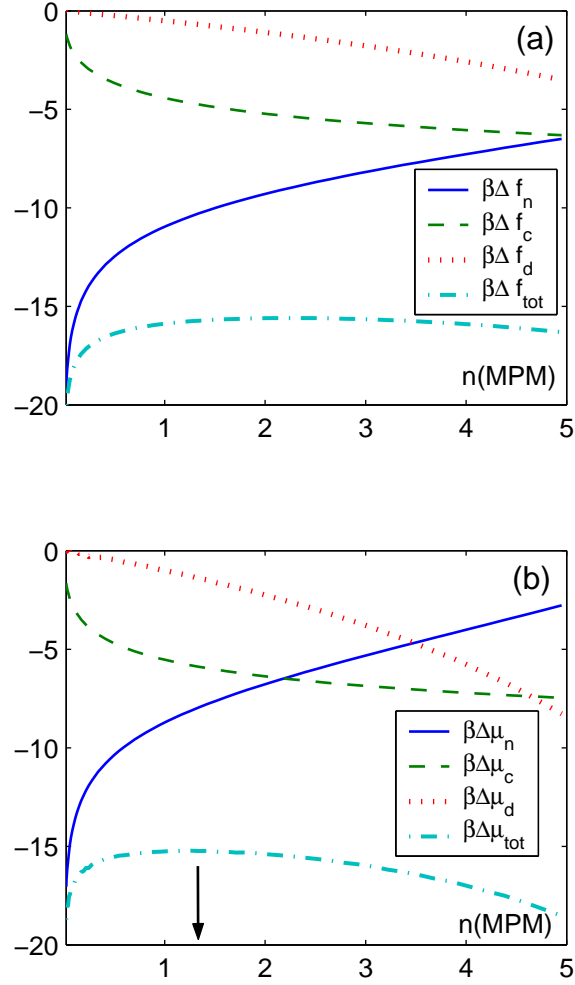} \caption{The dimensionless contributions to the excess free energy (a), to the excess chemical potential
(b) at $T=-70^0$C. The nonpolar contributions are depicted by solid lines, the dispersion ones by the dashed lines, the electrostatic ones by
the dotted lines, and the total change by the dashed-dotted lines. The arrow indicates the critical concentration $n_{s}$ at which the
compressibility is equal to infinity. All other conditions are the same as that in Fig. \ref{fig1}.} \label{fig9}
\end{figure}

\begin{figure}[tbp]
\includegraphics{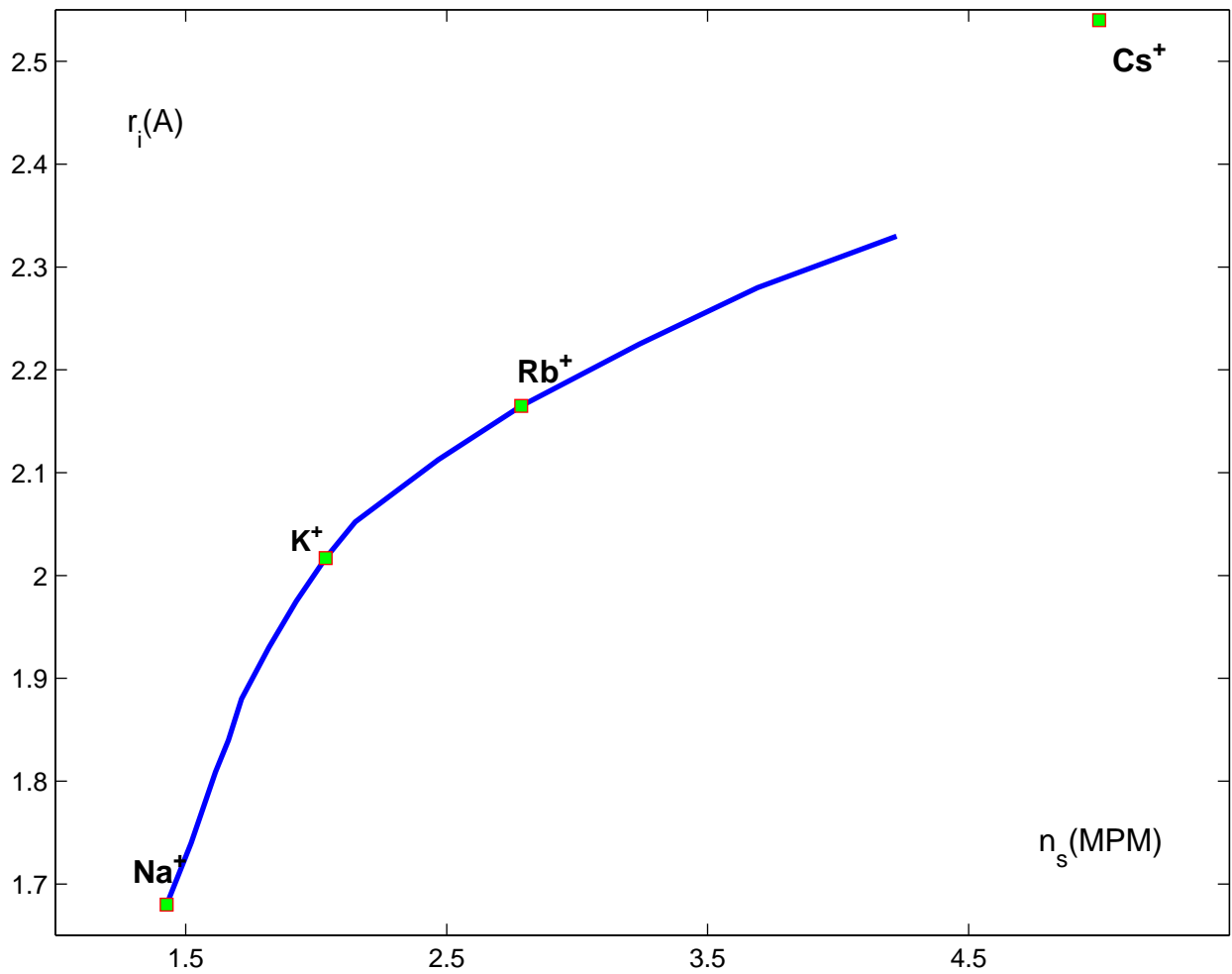}
\caption{Influence of the ion radius on the locus of the spinodal point $n_{s}$. The symbols corresponds to the van der Waals radii of the
relevant ions, derived from \cite{jorg}. All other conditions corresponds to that in Fig. \ref{fig1}.} \label{fig11}
\end{figure}

\begin{figure}[tbp]
\includegraphics{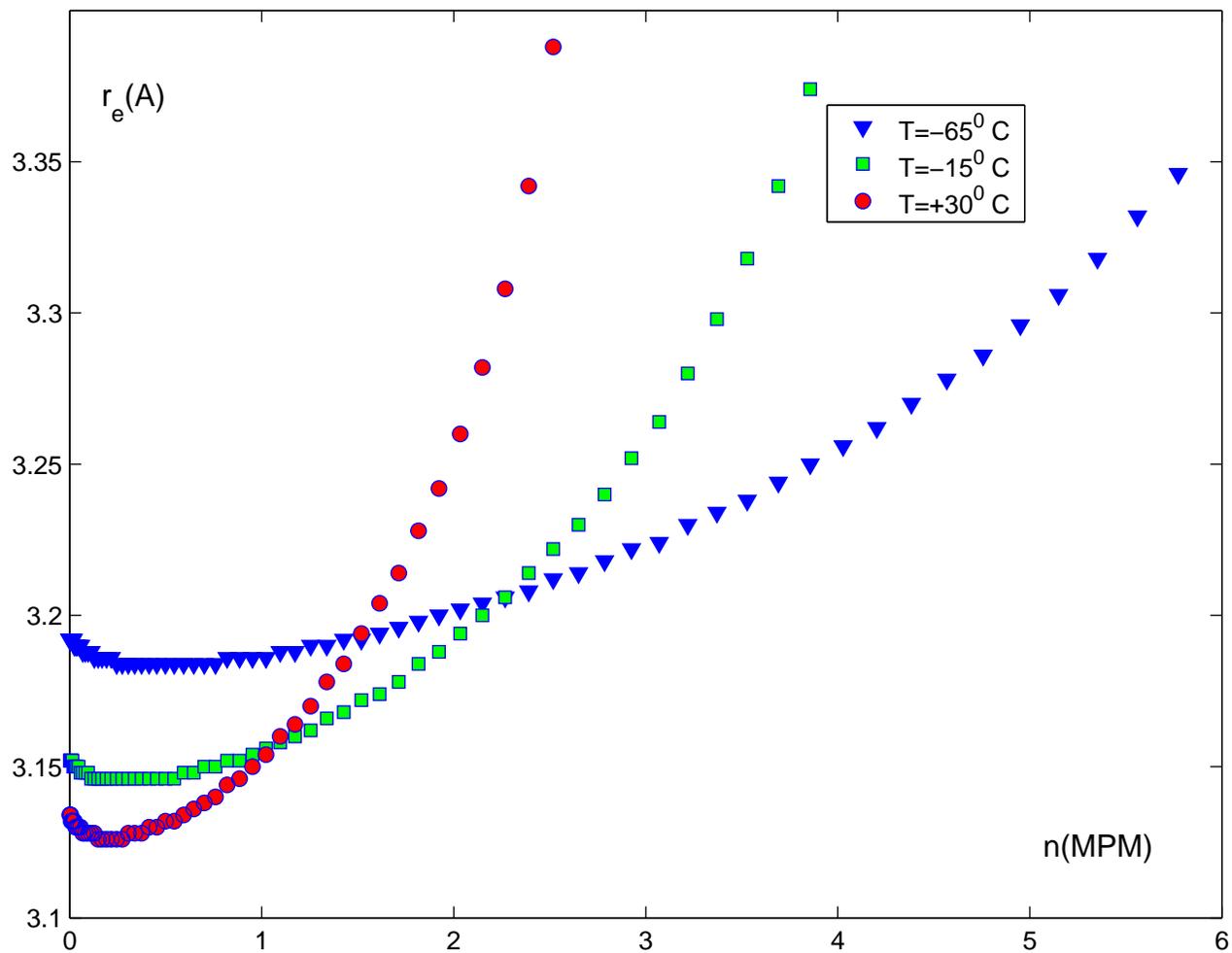}
\caption{Concentration dependence of the cavity radius $r_e(n)$ at various temperatures.} \label{fig10}
\end{figure}

\begin{figure}[tbp]
\includegraphics{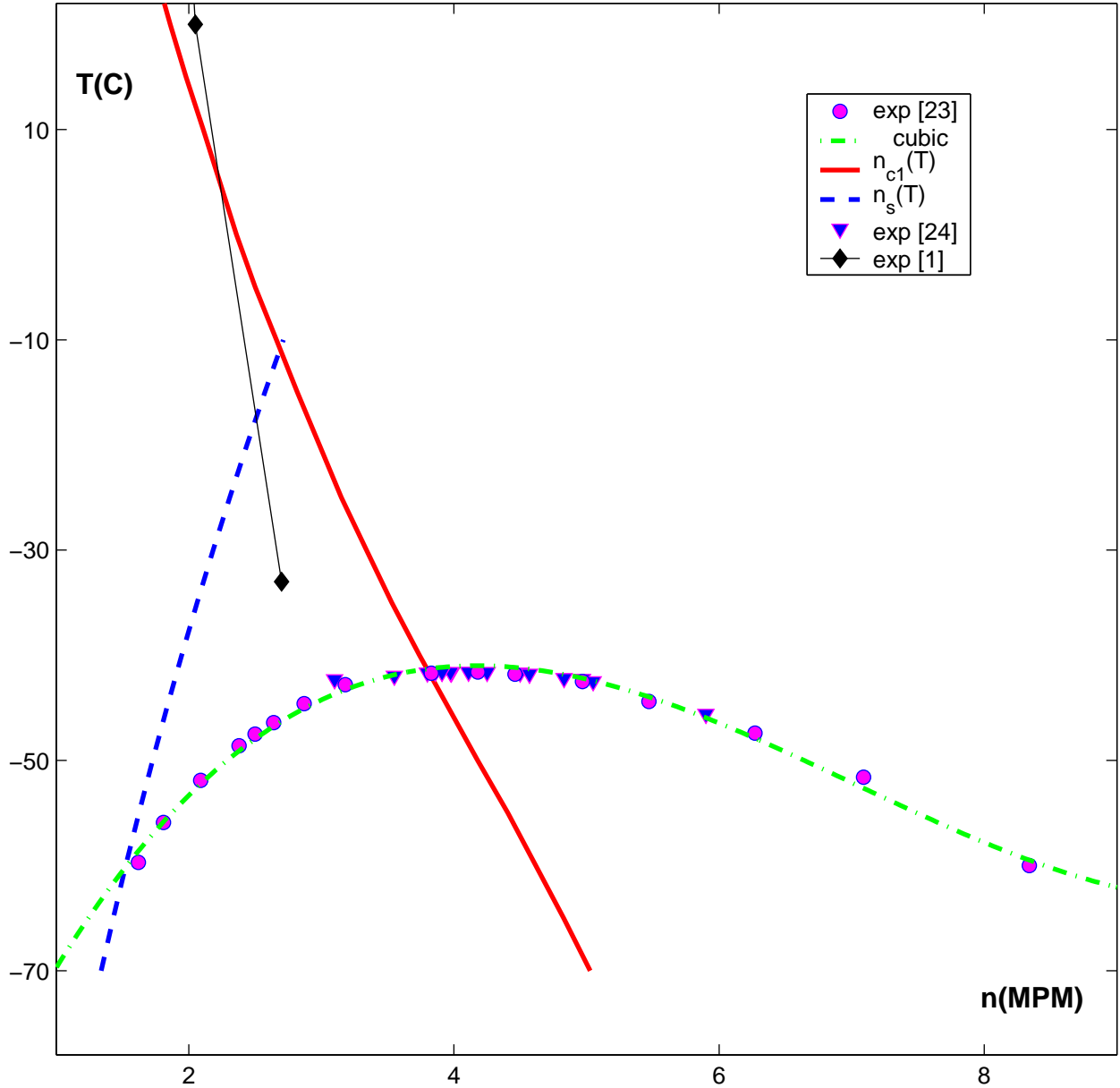}
\caption{ Phase diagram of Na-NH$_{3}$ solution. The experimental data on the locus of the phase separation are indicated by squares
\cite{crauss} and triangles \cite{CHIEUX}, respectively, while the dashed-dotted line indicates the cubic interpolation of these data. The
diamonds show the change in sign of the derivative of the conductivity coefficient $d\sigma /dT$, which is used to estimate the locus of the MNM
transition \cite{thompson}. The dashed curve corresponds to the low bond of the spinodal. The solid curve depicts our calculations of the locus
of the polarization catastrophe. The bottom of the figure corresponds to the solidification of ammonia.} \label{fig12}
\end{figure}


\begin{thebibliography}{99}
\bibitem{MPM}  Mole fraction is calculated as a ratio of moles of the metal
and that of the solution, 1(MPM) $\approx 2\cdot 10^{20}$ cm$^{-3}$ in MAS at $T=-70^0$C.
\end{thebibliography}
\end{document}